\documentclass[trackchanges]{aastex701}

\usepackage{amsmath}
\usepackage{amssymb}
\usepackage{xcolor}

\begin{document}

\title{The \textit{Swift}/UVOT Observation Campaign of Comet C/2013 US$_{10}$ (Catalina) and Near-nucleus UV Excess}

\author[orcid=0000-0001-6422-1038,sname=Shawn,gname=Oset]{Shawn Oset}
\affiliation{Department of Physics, Edmund C. Leach Science Center, Auburn University, Auburn, AL 36849, USA}
\email[show]{szo0032@auburn.edu}

\author[0000-0003-2399-5613]{Zexi Xing} 
\affiliation{Department of Physics, Edmund C. Leach Science Center, Auburn University, Auburn, AL 36849, USA}
\email{zzx0030@auburn.edu}

\author[0000-0003-2152-6987]{John Wm. Noonan}
\affiliation{Department of Physics, Edmund C. Leach Science Center, Auburn University, Auburn, AL 36849, USA}
\email{noonan@auburn.edu}

\author[0000-0003-2110-8152]{Steven J. Bromley} 
\affiliation{Department of Physics, Edmund C. Leach Science Center, Auburn University, Auburn, AL 36849, USA}
\email{szo0032@auburn.edu}

\author[0000-0002-2668-7248]{Dennis Bodewits}
\affiliation{Department of Physics, Edmund C. Leach Science Center, Auburn University, Auburn, AL 36849, USA}
\email{dennis@auburn.edu}

\correspondingauthor{Shawn Oset}

\begin{abstract}

We present results of a multi-wavelength extended monitoring campaign of dynamically new long-period comet C/2013 US$_{10}$ (Catalina), captured with the Ultraviolet and Optical Telescope (UVOT) on board the Swift spacecraft covering 5.8 AU inbound to 5 AU outbound.
The pipeline and methodology developed for these results is introduced, along with derived water production rates, active areas, and dust content proxy $A(\theta)f\rho$.
We describe a Bayesian solution to address the unknown spectral gradient $S'$ of the dust of C/2013 US$_{10}$ necessary for our continuum subtraction calculations.
The comet exhibited significant hyperactivity, with the water production rate approaching $2.5 \times 10^{29}$/s and a greatly elevated active area pre-perihelion, settling to approximately 50 km$^2$ post-perihelion.
We also report and quantify a near-nucleus UV excess beyond vectorial model predictions, a feature we have dubbed the Pale Blue Spot (PBS), and cover possible candidate species responsible.
When interpreted as OH (A-X) band fluorescence, the PBS may point to a secondary release mechanism producing slow-moving hydroxyl radicals that is active even beyond the water snow line.

\end{abstract}


\section{Introduction}\label{sec:intro}
Comets are among the most primordial remnants of Solar System formation, preserving a record of volatile ices and refractory dust from the proto-planetary disk.
As these bodies approach the Sun, sublimation of volatiles drive the development of a coma, offering a direct means to probe their chemical inventory \citep{AHearn_2011}.
The main component of the coma is water \citep{Biver_2024}, and therefore its storage and release are central to understanding cometary activity.
Quantifying H$_2$O production rates is then fundamental not only for insight into individual cometary activity but also to constraining the broader paradigms of volatile delivery and chemical evolution in the early solar system.
There are different methods to remotely detect water and its fragment species \citep[c.f.][]{Combi_2004, comets_ii_feldman_2004}, and near-UV emission from OH fragments produced by water photodissociation provides an accessible proxy for measuring cometary water production.

The Neil Gehrels \textit{Swift} Observatory (\textit{Swift})’s Ultraviolet/Optical Telescope (UVOT) offers a unique vantage for cometary studies  with its suite of broadband UV and optical filters \citep{Gehrels2004,Roming2000}.
Through a filter‐subtraction technique leveraging the UVW1 (centered near 2600 \AA) and V (centered near 5460 \AA) filters, one can isolate the OH emissions from the underlying dust continuum \citep{Bodewits_2011,Bodewits2023} as an indirect measure of the water production rate when combined with photochemical coma modeling.
Applying this methodology to images acquired by a tailored reduction pipeline enables measurement of H$_2$O release rates, active areas, and dust-to-gas ratios through the dust proxy $A(\theta)f\rho$.

The \textit{Swift} public data archive currently contains images of 44 unique comets composed of more than 5,000 observations, spanning almost two decades.
A large portion of the available comet images have been analyzed partially or not at all.
Comet C/2013 US$_{10}$ (Catalina) presents an especially compelling target: as a dynamically new or at least young comet \citep{Combi_2018}, its inbound activity offers a rare glimpse into volatile and dust characteristics before significant solar processing.
Its favorable apparition in late 2014–late 2016 allowed for a dense UVOT monitoring campaign spanning both pre‐ and post‐perihelion geometries.
In this study, we describe the custom pipeline deployed for UVOT images, outline our statistical approach to filter-subtraction when the dust is not characterized, and report H$_2$O production rates, active areas, and dust content for C/2013 US$_{10}$.

\section{Observations}
The Neil Gehrels \textit{Swift} Observatory \citep{Gehrels2004} is a multi‐wavelength, rapid‐slewing spacecraft originally designed for the prompt detection and follow‐up of gamma‐ray bursts (GRBs), launched in November 2004.
Its UltraViolet/Optical Telescope (UVOT) is a 30 cm modified Ritchey–Chrétien telescope that provides imaging and low‐resolution spectroscopy over the 1600–8000 \AA~range, enabling ultraviolet and optical observations of transient targets \citep{Roming2000} with a 17 x 17\arcmin\  field of view  and a plate scale upper limit of 0.502\arcsec\  pixel\(^{-1}\).

\begin{figure*}
    \centering
    \includegraphics[width=0.8\linewidth]{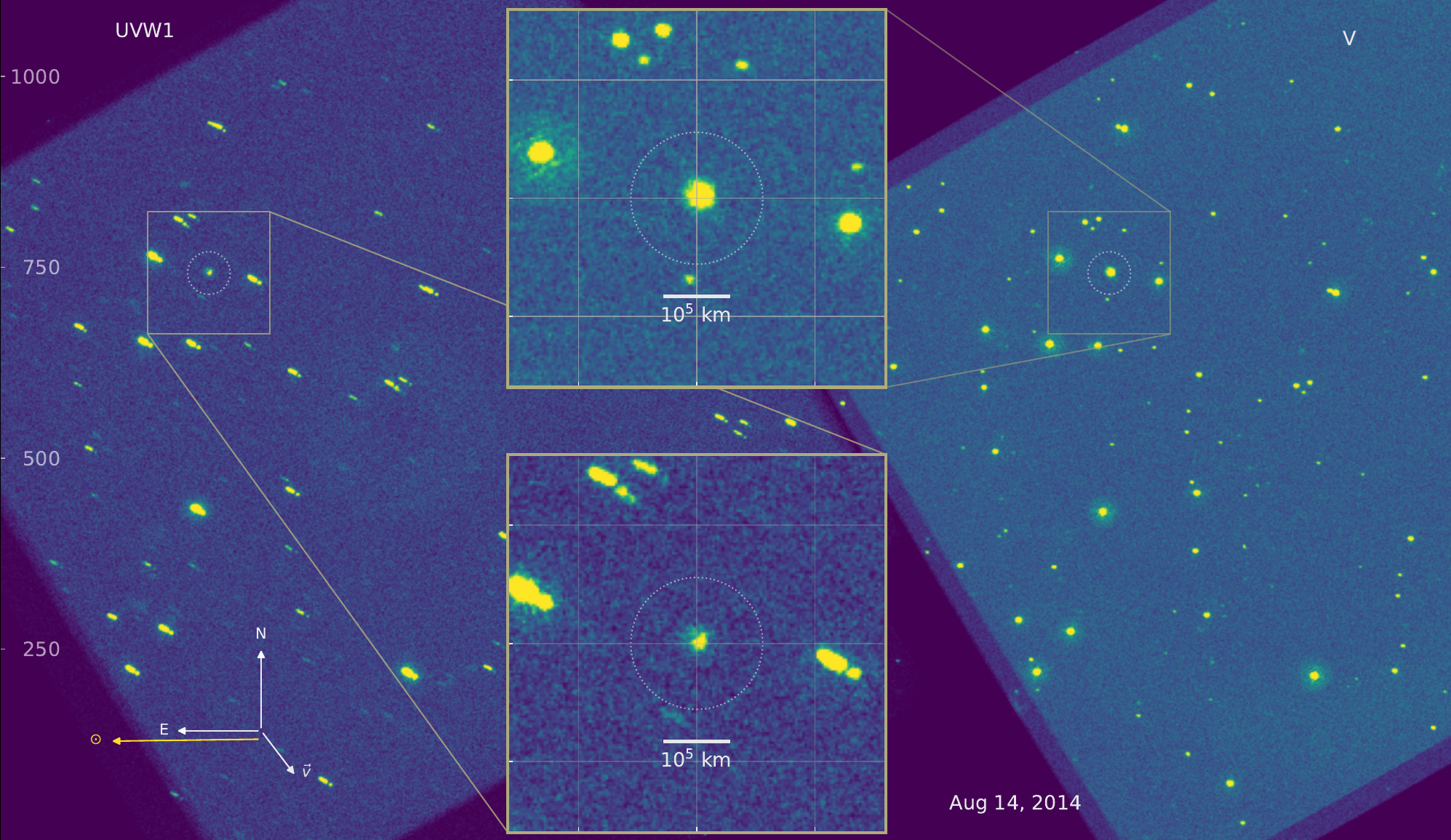}
    \centering
    \includegraphics[width=0.8\linewidth]{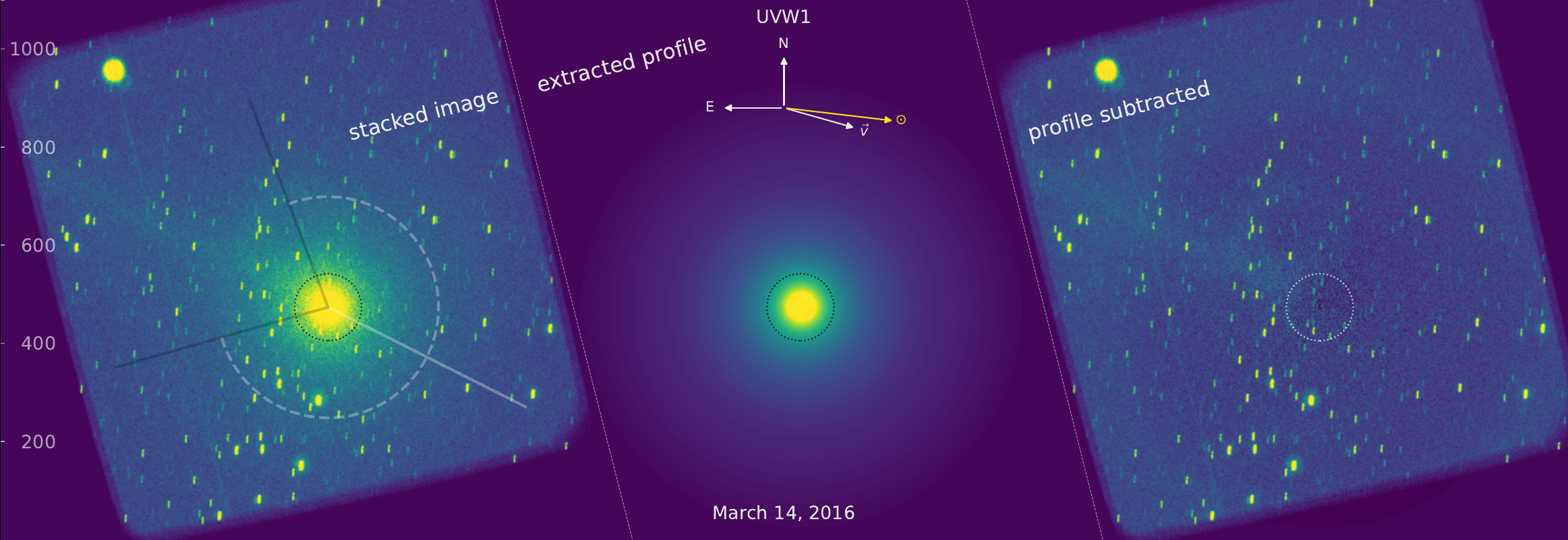}
    \includegraphics[width=0.8\linewidth]{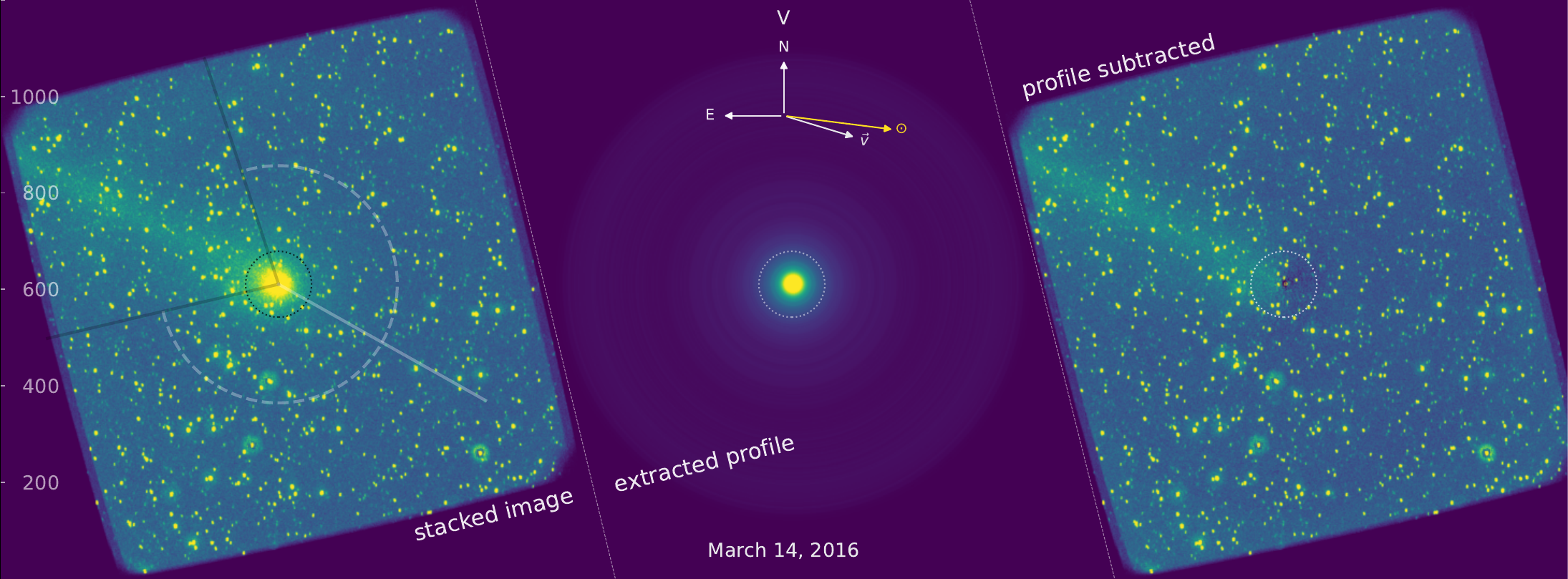}
    \caption{Top: \textit{Swift}/UVOT sum-stacked, background-subtracted images of C/2013 US$_{10}$, epoch 1 at 5.8 AU inbound.  Top left: UVW1, top right: V.  Color scales are independent and chosen for viewing.  Projected aperture of radius 10$^5$ km is in dotted white.  Lower inset of top image: UVW1, upper inset: V.  Inset grid lines are 50 pixels ($1.79 \times 10^5$ km) apart.  Both filters exhibit a small deviation of comet center from predicted JPL/Horizons ephemerides. Middle image: \textit{Swift}/UVOT sum-stacked, background-subtracted images of C/2013 US$_{10}$, epoch 9 at $r_h = 2.15$ AU, $\Delta = 2.09$ AU outbound on March 14, 2016 in UVW1 filter. Left: stacked images with profile extraction region highlighted.  Dotted circles mark projected 10$^5$ km aperture.  Solid white line marks the middle of the extraction region, while black solid lines mark the edges, and dashed white line shows the angular extent.   Middle: Extracted median profile expanded to two dimensions ($I_s$).  Right: stacked images with extracted profile subtracted ($I_a$). Left and right image color scales are fixed per filter, chosen for viewing.  The middle profile color scales chosen to illustrate extracted portion of the coma. In addition to star removal, the UV reflection donut in the UVW1 filter is left behind by profile extraction.  Bottom: V filter images for epoch 9, similar to middle.}
    \label{fig:showcase}
\end{figure*}

Comet C/2013~US$_{10}$ (Catalina) was observed over 12 epochs by \textit{Swift}/UVOT between August 2014 and November 2016 ranging from 5.8 AU inbound to 5.0 AU outbound, using the V (central \(\lambda = 5468\) \AA, FWHM 769 \AA) and UVW1 (central \(\lambda = 2600\) \AA, FWHM 693 \AA) filters.
The observing log is summarized in Table~\ref{table:observation_log} and a representative sample of the images is shown in Figure~\ref{fig:showcase}.
The V filter provides images of the dust in a spectral region with expected minor contamination (e.g., C$_2$, NH$_2$).
The UVW1 filter captures OH A$^2\Sigma^+$ - X$^2\Pi$ spectral emission features and dust, with possible contaminants discussed in Section \ref{sec:pbs}.
The observational campaign sampled the comet across a broad range of heliocentric distances, subject to \textit{Swift}’s $>$50\textdegree\  solar‐elongation requirement and onboard tracking limitations.

\startlongtable
\begin{deluxetable}{cccccccccc}\label{table:observation_log}
    \tablewidth{0pt}
    \tablecaption{\textit{Swift}/UVOT observation log for C/2013 US$_{10}$}
    \tablehead{\colhead{Epoch} & \colhead{Mid Time} & \colhead{$r_h$} & \colhead{$\dot{r_h}$} & \colhead{$\Delta$} & \colhead{S-T-O} & \colhead{T - T$_p$} & \colhead{Filter} & \colhead{Images} & \colhead{Exposure Time} \\ & (yyyy-mm-dd) & (AU) & (km $s^{-1}$) & (AU) & ($^\circ$) & (days) & & & (s)}
    \startdata
    1 & 2014-08-14 & -5.83 & -16.2 & 4.94 & 5.18 & -458.5 & UVW1 & 12 & 2075\\
    & & & & & & & V & 4 &  774\\
    2 & 2014-11-05 & -5.03 & -17.2 & 4.65 & 10.88 & -375.6 & UVW1 & 12 & 2050\\
    & & & & & & & V & 4 &  771\\
    3 & 2014-12-21 & -4.57 & -17.8 & 4.97 & 10.82 & -330.5 & UVW1 & 12 & 2051\\
    & & & & & & & V & 4 &  775\\
    4 & 2015-04-28 & -3.16 & -20.4 & 3.62 & 15.16 & -201.7 & UVW1 & 12 & 2054\\
    & & & & & & & V & 6 &  888\\
    5 & 2015-06-19 & -2.52 & -21.8 & 2.13 & 23.32 & -149.3 & UVW1 & 11 & 1995\\
    & & & & & & & V & 6 &  888\\
    6 & 2015-08-11 & -1.83 & -23.1 & 1.09 & 28.30 & -96.2 & UVW1 & 12 & 1991\\
    & & & & & & & V & 4 &  775\\
    7 & 2015-09-01 & -1.56 & -23.2 & 1.21 & 40.35 & -75.8 & UVW1 & 16 & 2884\\
    & & & & & & & V & 4 &  773\\
    8 & 2016-02-11 & 1.72 & 23.2 & 1.13 & 32.89 & 87.0 & UVW1 & 15 & 2727\\
    & & & & & & & V & 4 &  772\\
    9 & 2016-03-14 & 2.15 & 22.6 & 2.09 & 27.09 & 119.1 & UVW1 & 10 & 1938\\
    & & & & & & & V & 4 &  775\\
    10 & 2016-04-10 & 2.50 & 21.8 & 2.89 & 19.78 & 146.3 & UVW1 & 12 & 2051\\
    & & & & & & & V & 4 &  773\\
    11 & 2016-08-19 & 4.02 & 18.7 & 4.43 & 12.56 & 277.3 & UVW1 & 21 & 3634\\
    & & & & & & & V & 6 & 1159\\
    12 & 2016-11-24 & 5.03 & 17.2 & 4.20 & 6.65 & 374.3 & UVW1 & 18 & 3076\\
    & & & & & & & V & 6 &  866\\
    \enddata
\end{deluxetable}

\section{Pipeline Overview}
\subsection{Justification}
Piecing together published \textit{Swift} results for a broad-context picture suffers from differences in methodology, estimated values, data processing methods, and modeling assumptions in the individual analyses.
The pipeline [github link removed for anonymous review] described in this work is aimed at remedying  both the inherent spread of values introduced by separate analyses and the un-analyzed stock of data.
A related concern, especially when results between teams disagree, is reproducibility.
The pipeline described in this work tracks all intermediate and final output products for a given comet with a simple build system that ensures consistency and reproducibility when any input parameters change.
This enables a systematic, multi-decade analysis allowing direct comparison across comets by the uniformity of methodology.

\subsection{Data Preparation}\label{sec:data_prep}
\textit{Swift}/UVOT images of C/2013 US10 are prepared for analysis by compiling calibrated level 2 observations, available for download on the public \textit{Swift} archive, into a standardized observation database.
FITS metadata are extracted and target names standardized across inconsistent archive naming conventions.
Ephemerides from JPL Horizons \citep{Giorgini_1996,Ginsburg_2019} are used to determine the comet position in each exposure for subsequent alignment.
Before stacking, individual images are corrected for photon coincidence loss \citep{Poole_2007} using a PSF-based, flux-preserving convolution to estimate and correct local count-rate suppression, applied over the entire image.
An additional correction for time-dependent sensitivity loss is applied to the count rates to account for the roughly 1\% count rate loss per year \citep{Breeveld_2011}.
Event-mode images, taken at a 0.502\arcsec/pixel$^{-1}$ scale, are 2x2 re-binned to match the image-mode pixel scale of 1.004\arcsec/pixel$^{-1}$ and divided into short time slices and stacked to correct for comet motion during an exposure.
Corrected exposures are then visually screened for artifacts, aligned on the comet nucleus, and co-added by filter and epoch to produce stacked images for radial-profile extraction and aperture photometry.
The stacked images are then background subtracted, though in some cases subtraction is skipped because the coma can occupy the entire field of view.

\subsection{Epoch Analysis}\label{sec:epoch_analysis}
For each epoch, the OH emission in the UVW1 filter is isolated from dust-reflected solar light through continuum subtraction by scaling the V filter dust image to predict a dust count rate in the UVW1 filter.
The V filter scaling requires knowledge of the spectral gradient $S'$, the wavelength-dependent reflectivity of the cometary dust, that is unknown without supporting measurements from other instruments.
A Bayesian treatment of the unknown spectral gradient $S'$ is applied to derive OH aperture photometry and OH radial profiles.
For distant observations where profiles are not recoverable, water production rates are taken from OH aperture photometry by forward-modeling the coma with a vectorial model \citep{Festou_1981} to match OH molecular count within the aperture.
When OH profiles are available, we also derive water production rates through directly fitting a vectorial model to the OH column density.
The active area is then estimated using these water production rates using a sublimation model \citep{CowanAHearn1979}.
The dust content proxy $A(\theta)f\rho$ is computed using both aperture photometry and radial profiles taken from the V filter.

\section{Methods}

\subsection{Continuum Subtraction}\label{sec:continuum_subtraction}
To evaluate the dust contribution to the UVW1 filter, a solar spectral model \citep{Colina_1996} is convolved with the UVW1 and V filter effective areas to obtain the relative dust contribution $\beta(S')$ to each filter as a function of the spectral gradient $S'$.
We then calculate the residual OH flux emission $F_{OH}$ with
\begin{equation}\label{eq:count_rate_to_flux}
    F_{OH}(S') = \alpha \cdot \left( \dot{N}_{uvw1} - \beta(S') \cdot \dot{N}_{v}
 \right)
\end{equation}
where $\dot{N}$ is the photon count rate observed in a given filter and $F_{OH}$ is the redness-dependent OH flux derived from the respective count rates.
The count-rate-to-flux factor $\alpha$ is calculated by convolving the UVW1 filter with an OH spectral model \citep{Bodewits_2019} covering the A-X fluorescence bands centered around 3085\AA.
The number of emitting OH molecules in the field of view under consideration is then
\begin{equation}\label{eq:flux_oh_to_n_oh}
    N_{OH}(S') = \frac{4 \pi \Delta^2 F_{OH}(S')}{g_{OH}}
\end{equation}
where $g_{OH}$ is the gas fluorescence efficiency, scaled by $r_h^{-2}$, taken from the OH fluorescence emission model in \citet{Schleicher_AHearn_1988} to account for the heliocentric velocity dependent Swings effect, which incorporates the 0-0, 1-0, 1-1, and 2-2 transition bands.
In the case of aperture photometry, the OH number $N_{OH}(S')$ in the aperture is matched to a vectorial model to derive the water production Q(H$_2$O, $S'$).
For extracted radial flux profiles, Equation \ref{eq:flux_oh_to_n_oh} is applied pixel-wise to derive the column density as a function of radius and fitted with a vectorial model to find the resulting water production [see Section \ref{sec:vectorial_model}].

Assuming that contamination from other species in the UVW1 filter is not significant, $S'$ could in principle be constrained by searching for the value that produces the best vectorial fit.
In practice, however, this procedure has severe limitations.
The first is that the comet must not be too distant so that surface brightness profiles can be reliably extracted and fitted.
The second is that the outer coma contains little dust and therefore struggles to constrain the dust color, while the inner coma column density is often not well described by the vectorial model [see Section \ref{sec:near_nucleus_uv_excess}].
We therefore treat $S'$ as an uncertain parameter and describe our methodology to address this uncertainty in Section \ref{sec:dust_reddening}.

\subsection{Dust Reddening}\label{sec:dust_reddening}
\subsubsection{Reddening Translation}
The assumed value of dust reddening $S'$ is pivotal to estimating water production rates in the continuum subtraction method.
Measurements of a comet's dust reddening are not always available, and those that exist are likely not in the wavelength range utilized by the UVW1/V filter pair. To complicate the matter further, dust color is not necessarily a static property of a comet \citep[e.g., 67P][]{Kwon_2022}.
This is of particular importance because the dust reflectance, and therefore $S'$, is a function of wavelength.
To maximize the amount of relevant data, we may translate a spectral gradient $S'$ observed between filters with effective wavelengths $\lambda_1, \lambda_2$ to what the spectral gradient $S'_t$ would be if measured between filters $\lambda_3, \lambda_4$ with an assumption that the reflectance stays linear across the wavelength range translated over.
Starting with a linear reflectance assumption $S(\lambda) \propto m \lambda + b$ and adopting $\hat{\lambda}_{2,1} = (\lambda_2 + \lambda_1)/2$ as the mid wavelength of the filter pair:
\begin{align}
    S' &= \frac{\gamma}{\Delta \lambda_{2,1}} \frac{m \lambda_2 + b - m \lambda_1 - b}{m \lambda_2 + b + m \lambda_1 + b} \notag \\
    &= \frac{\gamma}{\Delta \lambda_{2,1}} \frac{m \Delta \lambda_{2,1}}{2m\hat{\lambda}_{2,1} + 2b}
\end{align}
where $\gamma = 2 \cdot 100\% \cdot 1000$/\AA\ and $\Delta \lambda_{2,1}$ is the difference in effective wavelength (as defined in \citealt{Koornneef_1986}) of the two filters.
Repeating this expression for $S'$ over the filter pair $\lambda_3, \lambda_4$, while retaining the same linear reflectance relation (i.e., keeping $m/b$ fixed), results in the translated spectral gradient in the new $\lambda_3, \lambda_4$ wavelength range:
\begin{equation}\label{eq:reddening_translation}
    S'_t = \frac{\gamma S'}{\gamma - 2 S' \left(\hat{\lambda}_{4,3} - \hat{\lambda}_{2,1}\right)}
\end{equation}
 so that the translated spectral gradient becomes a function of the middle wavelengths of the respective filter pairs.
This approach was previously applied by \citet{xing2025_3i}, where the mean translated spectral gradient agreed with a direct measurement over the corresponding wavelength interval, supporting the validity of the method.

This translation formula suffers when used to translate over large wavelength intervals due to the observed $S'$ flattening toward the low optical and infrared regions and steepening in the UV \citep{Jewitt_Meech_1986, Jewitt_2015, Hainaut_2012, Kulyk_2018}.
The overall effect is that when translating longer-wavelength spectral gradients down to the near-UV, there will be a significant, systematic underestimation of the predicted redness at the UVW1/V filter pair interval.
The systematic nature still lends itself to rough comparison between measurements reported in the literature.

\subsubsection{Reddening Statistics of Long-period Comets (LPCs)}\label{sec:lpc_stats}
It is often the case that no contemporaneous measurements of $S'$ exist for a given comet, either in the wavelength range of the UVW1/V filter pair, or outside of it where the translation formula in Equation \ref{eq:reddening_translation} could be applied to obtain a rough estimate of the dust reddening.
The dust redness is a function of its physical properties, which can evolve in time and vary intrinsically between comets, so adopting an arbitrary single value of $S'$ will distort any dependent results, and the extent of the dust-related error would be unknown.

The MBOSS database \citep{Hainaut_2012} contains color measurements organized by dynamical class, with sampling $10 < N < 40$ for LPCs in various wavelength ranges.
Although the error is large, additional data from \citet{Solontoi_2012}, obtained with Sloan Digital Sky Survey (SDSS) filters, show a rough trend of decreasing spectral gradient with increasing wavelength, based on comets captured incidentally in the survey.
An additional color survey of \citet{Jewitt_2015} presents 32 observations of LPCs with sample sizes similar to the MBOSS database with a more constrained filter selection.
A summary of spectral gradient statistics is given in Table \ref{table:lpc_spectral_gradients}, including the translated spectral gradient as seen by the UVW1/V filter pair in $S'_t$.

\begin{deluxetable}{cccccccc}
    \tablewidth{0pt}
    \tablecaption{Spectral gradients of various LPCs \label{table:lpc_spectral_gradients}}
    \tablehead{\colhead{Source} & \colhead{Filters} & \colhead{Color mag} & \colhead{$\hat{\lambda}_{2,1}$} & \colhead{$S'$} & \colhead{$\hat{\lambda}_{4,3}$} & \colhead{$S'_t$} & N \\ & & & \AA & \%/1000\AA & \AA & \%/1000\AA &}
    \startdata
    MBOSS & B - V & 0.79 $\pm$ 0.07 & 4904.5 & 11.59 $\pm$ 5.9 & 4381.8 & $12.34^{+6.9}_{-6.5}$ & 33\\
    Hainaut et al. 2012      & B - R & 1.25 $\pm$ 0.09 & 5384.0 & 10.98 $\pm$ 4.00 &        & $12.34^{+5.26}_{-4.85}$ & 32\\
          & V - R & 0.46 $\pm$ 0.06 & 5927.5 & 9.98 $\pm$ 5.75 &        & $11.80^{+8.97}_{-7.28}$ & 36\\
          & B - I & 1.64 $\pm$ 0.11 & 6170.5 & 7.24 $\pm$ 2.75 &        & $8.31^{+3.82}_{-3.46}$ & 11\\
          & V - I & 0.91 $\pm$ 0.13 & 6714.0 & 7.58 $\pm$ 4.68 &        & $9.21^{+7.91}_{-6.13}$ & 15\\
          & R - I & 0.44 $\pm$ 0.07 & 7193.5 & 5.56 $\pm$ 4.09 &        & $6.59^{+6.64}_{-5.06}$ & 15\\
    \hline
     SDSS & u - g & 1.51 $\pm$ 0.38 & 4156.4 & 6.00 $\pm$ 28.07 &        & $5.92^{+25.56}_{-29.34}$ & 9\\
    Solontoi et al. 2012& g - r & 0.55 $\pm$ 0.11 & 5500.5 & 7.14 $\pm$ 6.77 &        & $7.76^{+8.68}_{-7.40}$ & 9\\
    & r - i & 0.23 $\pm$ 0.09 & 6928.0 & 8.07 $\pm$ 6.21 &        & $10.15^{+12.24}_{-8.22}$ & 9\\
    & i - z & 0.06 $\pm$ 0.29 & 8361.0 & 1.74 $\pm$ 18.14 &        & $1.87^{+92.55}_{-11.80}$ & 9\\
    \hline
    Jewitt et al. 2015 & B - V & 0.78 $\pm$ 0.02 & 4904.5 & 10.75 $\pm$ 1.69 &        & $11.39^{+1.91}_{-1.88}$ & 26\\
    & B - R & 1.24 $\pm$ 0.02 & 5384.0 & 10.54 $\pm$ 0.89 &        & $11.78^{+1.12}_{-1.10}$ & 25\\
    & V - R & 0.47 $\pm$ 0.02 & 5927.5 & 10.94 $\pm$ 1.91 &        & $13.16^{+2.88}_{-2.68}$ & 25\\
    & R - I & 0.42 $\pm$ 0.03 & 7193.5 & 4.39 $\pm$ 1.75 &        & $5.01^{2.42}_{-2.16}$ & 9\\
    \enddata
    \tablecomments{Values computed from \citet{Solontoi_2012} using LPCs only.  N indicates number of comets sampled.  Quantity $\hat{\lambda}_{x,y}$ describes the middle wavelength between filters $x$ (upper wavelength) and $y$ (lower wavelength), with wavelengths as defined in the Johnson-Cousins or Sloan filter sets.}
\end{deluxetable}

The trend of increased reddening at lower wavelengths, combined with the positioning of the UVW1 filter in the near-UV, present a challenge when applying this translation to find $S'_t$.
Reddening values as high as 60 to 80\%/1000\AA\ were recorded by the IUE and HST near 2950 \AA~for several comets \citep{Feldman_AHearn_1985}, while the average translated reddening in the B - V (4400 to 5500 \AA) range for LPCs is 12\%/1000\AA\ assuming $(B - V)_{\odot} = 0.653$ \citep{Ramirez_2012}.
The UVW1 filter lies in the UV wavelength region exhibiting extreme reddening.
It should be expected, when paired with the V filter, that together they capture some intermediate $S'$ value under the linear reddening model: smaller than the UV reddening but larger than the predicted $S'_t$ reddening implied by the optical wavelengths surveys in Table \ref{table:lpc_spectral_gradients}.

\subsubsection{Statistical Approach to Reddening}\label{sec:stat_approach_reddening}
As the data typically does not allow for best-fit OH profiles to determine the dust redness (see Section \ref{sec:continuum_subtraction}), and contemporaneous $S'$ measurements at other wavelengths are often unavailable, we therefore assume that the typical long-period comet comes from a prior distribution of rednesses, $p(S')$, as seen in our filter pair and adopt a Bayesian framework in which $S'$ is treated as a random variable.
We adopt a prior probability density function proportional to a normal distribution $N_{S'}$ with mean redness $\mu_{S'}$ and standard deviation $\sigma_{S'}$ 
\begin{equation}\label{eq:bayes_dust_prior}
    p(S') \propto N_{S'}(\mu_{S'} = 20 \%/1000\ \text{\AA}\ , \sigma_{S'} = 3\%/1000\ \text{\AA})
\end{equation}
so that when randomly sampling the population of LPCs, on average the observed reddening as seen by the UVW1 and V filter pair would be $\mu_{S'}$ with standard deviation $\sigma_{S'}$.
The motivation for the particular values of $\mu_{S'}$ and $\sigma_{S'}$ is discussed in Section \ref{sec:lpc_stats}, but are not definitive: while informed by the available data, the values are still rough estimates that may be revised in future work.
The pipeline uses this prior as a standard for best-estimate values that are downstream of continuum subtraction: flux, water production rates, and active areas.

With our prior $p(S')$ we may now address the unknown dust reddening $S'$ statistically.
We could take the prior $p(S')$ and derive an expected OH count $N_{OH}$ by marginalizing over $S'$ as a nuisance parameter:
\begin{equation}
    \left<N_{OH}\right> = \int p(S') N_{OH}(S') dS'
\end{equation}
but during continuum removal, $F_{OH}$ from Eq. \ref{eq:count_rate_to_flux} and therefore $N_{OH}$ is not guaranteed to be positive, especially at more blue spectral gradients $(S' < 0)$.
Blue colors represent subtracting larger and larger quantities of dust seen in the V filter from the UVW1 signal, and may result in negative values due to over-subtraction.
Artificially low expectation values for $N_{OH}$ will occur if we include un-physical negative-valued results in the above calculation indiscriminately.

Instead, we apply a condition $\mathcal{A}$ to our prior that the outcome of continuum subtraction must be positive, so that we have a truncated prior $p(S' | \mathcal{A})$ that only samples physical outcomes.
To achieve this, we move to a quantized space for $S'$ and define $\left\{S'_p\right\}$ and $\left\{S'_u\right\}$ as the set of values of $S'$ that produce physical and unphysical outcomes, respectively.
Then the admissible prior mass is $P(\mathcal{A}) = \sum_{S' \in S'_p} p(S')$ and our truncated prior is normalized with
\begin{equation}
    p(S' | \mathcal{A}) = \frac{p(S')}{P(\mathcal{A})}, S' \in S'_p
\end{equation}
We define $\hat{r}$ as the rejected prior fraction, i.e., the prior mass assigned to spectral gradient values that yield unphysical results.
The rejected prior fraction quantifies the probability that our data suggest a non-detection of OH if the comet in question comes from our overall prior distribution $p(S')$ in equation \ref{eq:bayes_dust_prior}.

\begin{figure*}
    \centering
    \includegraphics[width=1.0\linewidth]{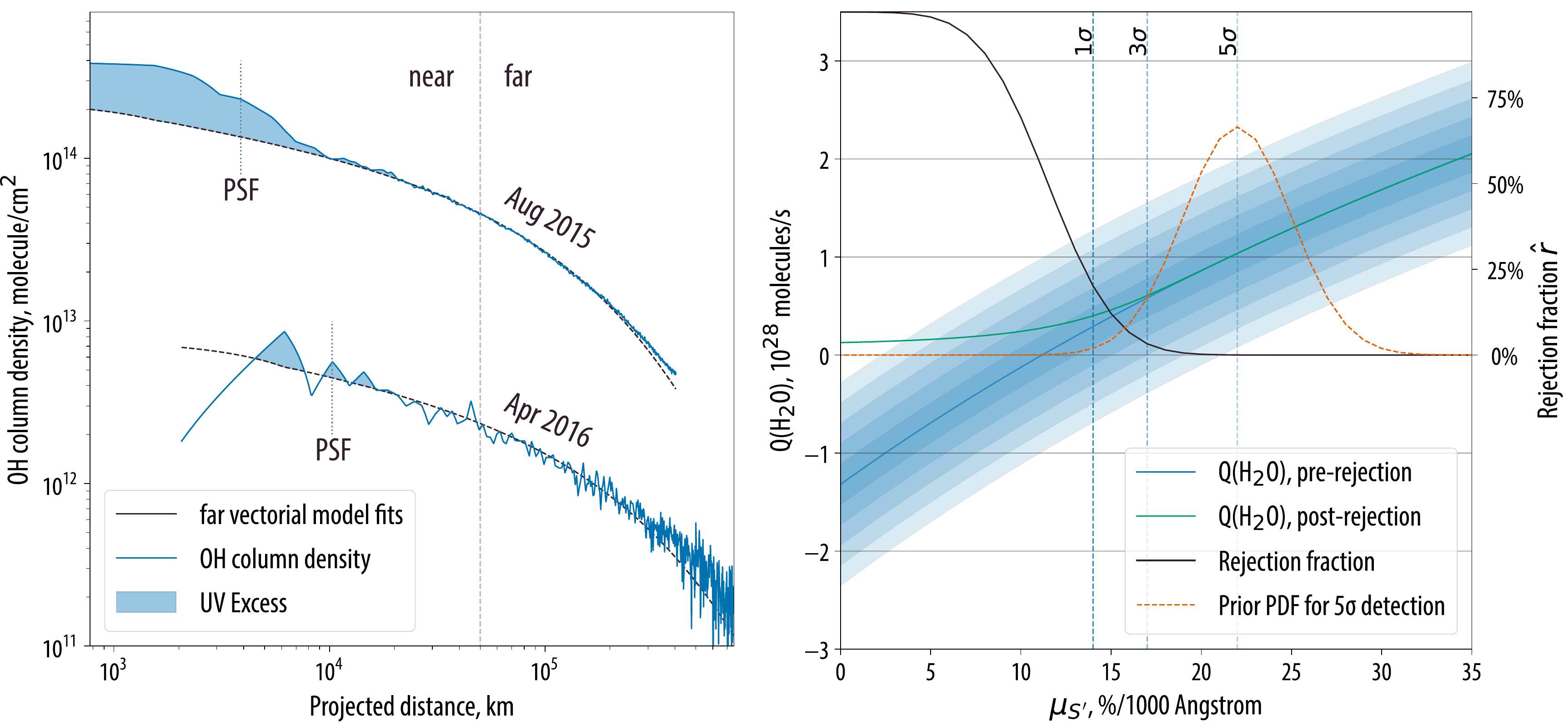}
    \caption{Left: Derived OH column densities and vectorial model fits for pre-perihelion epoch 6 and post-perihelion epoch 10 visits with OH UV excess highlighted.  Column density derived using the dust prior in Equation \ref{eq:bayes_dust_prior}. Division of near/far coma at $r = 5\times10^4$ km marked with gray dashed line [See Section \ref{sec:coma_split}].  Vectorial fits in black, based on the far region of the coma only.  Right: Water production rate from epoch 2 at 5.03 AU, derived from aperture photometry as a function of the dust prior PDF.  Solid blue and dark green curves show water production before and after rejecting unphysical sections of the dust prior, respectively, as a function of the mean of the dust prior.  Example dust prior PDF of $\mu_{S'} = 22\%/1000$~\AA~shown in dashed red, resulting in $5\sigma$ detection of OH. Rejection fraction shown in black, reaching nearly 100\% for this epoch if we believe the average comet color is gray. Vertical dashed lines: values of assumed $\mu_{S'}$ of 14\%, 17\%, and 22\%/1000 \AA\ that result in 1$\sigma$, 3$\sigma$, and 5$\sigma$ detections respectively.}
    \label{fig:oh_excess_and_aperture_photometry}
\end{figure*}

Using our truncated prior we calculate the average of possible OH detections on a grid of $S'$ values with equal grid spacing $\Delta S'$ as
\begin{equation}\label{eq:bayes_num_oh}
    \left<N_{OH}(\mathcal{A})\right> = \sum_{S'_i \in S'_p} p(S'_i | \mathcal{A}) \cdot N_{OH}(S'_i) \cdot \Delta S'
\end{equation}
so that $\left<N_{OH}(\mathcal{A})\right>$ represents the most likely value under condition $\mathcal{A}$.
In practice, $\left<N_{OH}\right>$ and $\left<N_{OH}(\mathcal{A})\right>$ are equivalent for a given prior unless the signal to noise ratio is less than $3$ (Figure \ref{fig:oh_excess_and_aperture_photometry}), where rejections become a significant fraction of the prior.
We therefore report only pre-rejection values in this work, but note the subtlety for future applications of the method.

To fully characterize the results of our data, we repeat the above process with a grid of Bayesian priors $\mu_{S'}$, with $\sigma_{S'}$ fixed, to find the dependence of the detection on our assumed prior as in Fig \ref{fig:oh_excess_and_aperture_photometry} after converting $\left<N_{OH}\right>$ to $\left<Q(H_2O)\right>$.
As new data about the color distribution of LPCs become available, we may revise our prior and take the associated updated value of water production.
In the case that the color of a comet is known, we can replace the prior $p(S')$ with a distribution that reflects the value and uncertainty of the measurement of $S'$ to derive water production rates, with the color error included, with no additional complexity.

\subsection{Aperture-based Water Production Rates}\label{sec:aperture_water_rates}
Aperture photometry is the only analysis path for distant observations, where aperture size effects may be significant due to extended water sources (See \ref{sec:discussion_water_production}).
We fix the the aperture size of the pipeline at a projected distance of $10^5$ km: larger than the PSF and large enough to incorporate the entire coma signal including distributed sources, but balanced against error growth, star contamination, and background gradient effects that grow with aperture size.
The image pair for epoch 1 at 5.8 AU inbound is shown in Fig.~\ref{fig:showcase} for context.
To convert the observed number of OH radicals within the aperture $\left<N_{OH}\right>$ into a water production rate, we forward model the OH column density using a vectorial model and match the modeled OH count within the same photometric aperture to the observed OH count.
This technique has been used to infer Q(H$_2$O) from OH observations because it provides a physically motivated description of the daughter species distribution resulting from photodissociation of water in the coma \citep{Combi_2004,Bodewits_2015} [see also Section \ref{sec:profile_water_rates}].
An example of aperture photometry results in Figure \ref{fig:oh_excess_and_aperture_photometry} shows the relationship between the Bayesian prior, the derived water production rate, and the significance of the detection.
Detailed results for each epoch can be found in Table \ref{table:aperture_photometry_summary}.

When converting the observed OH to water production rates, forward modeling with the vectorial model is not the only option.
We derive a mass-loss-equivalent (MLE) water production rate Q$_{MLE}$(H$_2$O), defined as the water production rate that, when used as the input for a vectorial model, correctly reproduces the measured mass loss rate $\dot{M}_{OH}$ of OH observed within an aperture.
The quantity $\dot{M}_{OH}$ is computed as $\left<N_{OH}\right>$ from Equation \ref{eq:bayes_num_oh} times the molecular mass of OH divided by the scaled lifetime, and is matched to the overall mass loss rate of OH in the coma of a vectorial model using Q$_{MLE}$.
Note that any OH outside the aperture is ignored, so a selected aperture must capture the entirety of the signal as shown in the top panel of Fig. \ref{fig:showcase}.
The Q$_{MLE}$ characterization is best applied in cases where the comet is too distant for OH column density profiles, or when additional sources of water or OH are active that conform poorly to coma models.
This puts Q$_{MLE}$(H$_2$O) in a similar role at large $r_h$ as the empirically-derived relation Q(H$_2$O) = $1.361 \cdot r_h^{-1/2} \cdot $ Q(OH) \citep{Schleicher_1998}, which was computed using observations mostly within 2.5~AU.

Although both approaches aim to translate observed OH molecular count $N_{OH}(S')$ into an equivalent water production rate, they can yield substantially different results under certain  conditions.
In particular, forward modeling can produce water production rates significantly larger than the directly inferred OH production, whereas $Q_{MLE}(H_2O)$ typically remains within the same order of magnitude as $Q(OH)$.
The implications of this difference are discussed further in Section~\ref{sec:discussion_water_production}.

\subsection{Profile-based Water Production Rates}\label{sec:profile_water_rates}

\subsubsection{The Vectorial Model and Profile Extraction}\label{sec:vectorial_model}
The vectorial model \citep{Festou_1981} predicts the distribution of fragment species in the coma after photo-dissociation of parent species that have sublimated off the nucleus.
It is an iteration of the Haser model \citep{Haser_1957, Haser_2020}, which envisions a constant flow of parent molecules traveling radially from the nucleus at constant speed, with fragments inheriting speed and direction of the parent.
Instead, the vectorial model introduces a vectorial `kick' to the fragments after photo-dissociation from excess photon energy.
For water dissociating into OH + H, the kick imparted to OH is approximately 1.05 km/s in a random direction \citep{Huebner2015ko}, scattering OH from water sources over a large volume.
As in the Haser model, the underlying assumptions of isotropic outflow of parents is the same.
Steady flow is not assumed in the time-dependent vectorial model offered in \verb|sbpy| \citep{Mommert2019}, but all calculations in this work assume that Q(H$_2$O) is constant during any given epoch.
The exposures that comprise an epoch are typically taken over the course of one to two days, so that water production, $r_h$, and associated $r_h$-adjusted quantities change negligibly.

The models provide a map from water production Q(H$_2$O) to OH column density profile, referred to in this work as $\Sigma_{OH}(\rho)$, with $\rho$ denoting the projected distance from the nucleus.
When mapping an observed $\Sigma_{OH}(\rho)$ to Q(H$_2$O) via fitting, the results are most applicable when the data meet the assumptions of the model: symmetric, steady outflow of parent molecules from the nucleus.
We therefore treat our UVOT images as being composed of an azimuthally symmetric component $I_s$ about the center of the nucleus, and asymmetric residue $I_a$, then take the V and UVW1 filter profiles for continuum subtraction from $I_s$. The extracted UVW1 and V profiles are combined pixel-wise using Equation \ref{eq:count_rate_to_flux} and a spatially-constant $S'$ to derive $\Sigma_{OH}(S', \rho)$.
An example of this process is shown in the bottom panels of Figure \ref{fig:showcase}, demonstrating a clean extraction of profiles from both filters as well as the disparate spatial scales of the OH and dust components of the coma.

\subsubsection{The Coma Split and Vectorial Fitting}\label{sec:coma_split}
In the inner coma, the $\rho^{-1}$ brightness profile of the dust dominates over the extended OH distribution.
Though the Bayesian prior of Equation \ref{eq:bayes_dust_prior} attempts to mitigate the impact of the dust and its unknown spectral gradient $S'$, we can further reduce its influence by avoiding the dusty inner region of the coma while fitting. 
We split the coma into a `near' and `far' region with a cutoff distance $\rho_c = 5\times10^4$ km and exclusively fit our extracted $\Sigma_{OH}(\rho, S')$ for $\rho > \rho_c$ to a vectorial model to obtain Q(H$_2$O, $S'$).
The value of $\rho_c$ was chosen to balance minimizing dust inclusion against using as much column density data for fitting as possible.
The derived values of Q(H$_2$O) are largely insensitive to $S'$ so that the error in  $\left<\text{Q(H}_2\text{O)}\right>$ after marginalization over $S'$ is dominated by fitting error, shown in Table \ref{table:production_rate_summary}.

While the dust is dilute in the far region of the coma, the V filter can also contain C$_2$ Swan band emissions that are spatially extended due to the long lifetime of C$_2$ in the coma \citep{Huebner2015ko}.
The production rate of C$_2$ is roughly half of the production rate of CN on average for comets \citep{Fink_1996}, though \citet{Opitom2016} indicates that for C/2013 US$_{10}$ the C$_2$ production is an order of magnitude lower than CN.
The fluorescence efficiency of CN is high but remains a minor contaminant as discussed in Section \ref{sec:contamination_cn}.
We treat C$_2$ contamination as insignificant in the V filter due to the combination of lower production rate and fluorescence efficiency.

\subsubsection{Near-nucleus UV Excess}\label{sec:near_nucleus_uv_excess}
Extrapolating the outer-coma vectorial model fits inward toward the nucleus sometimes reveals surprising discrepancies between the predicted and observed column densities in the inner coma. 
As shown in Fig.~\ref{fig:oh_excess_and_aperture_photometry}, the observed OH column density can exceed the vectorial model prediction close to the nucleus, up to a factor of two in the largest case.
This excess is prominent in the pre-perihelion observations but weakens rapidly after perihelion (See the quantity Q$^*$(OH) in Table \ref{table:production_rate_summary}), where the profile is generally well-described by a vectorial column-density profile.
We quantify this near-nucleus excess by integrating the difference between the radial column density profile and the vectorial model predictions within a radius $r'$ with
\begin{equation}\label{eq:blue_spot_integral}
    N^*(OH, S') = \int_{\theta=0}^{2\pi} \int_{\rho = 0}^{r'} (\Sigma_{OH}(S', \rho) - f_{vm}(\rho)) \cdot \rho \cdot d\rho \cdot d\theta
\end{equation}
where $f_{vm}(\rho)$ is the column density predicted by a fitted vectorial model, and $N^*(OH, S')$ is the inferred excess count of OH molecules.
Growth curves of $N^*(OH, S')$ indicate that the spatial scale of this feature is confined within $r' = 2 \times 10^4$ km when present for C/2013 US$_{10}$, and the implied OH production rates are similar in magnitude to OH production rates measured beyond 3 AU as seen in Fig.~\ref{fig:us10_lightcurve}.
It should be stressed that $N^*(OH, S')$ can also be converted back into flux via Equation \ref{eq:flux_oh_to_n_oh} to instead view this as a near-nucleus flux excess in the UVW1 filter, as it has not yet been established that this excess flux is the result of OH in the inner coma.

\subsection{Dust Content}
The quantity $A(\theta)f\rho$ provides a proxy for the total dust content within a photometric aperture \citep{AHearn_1984} when the dust is isotropic with constant outflow speed, where $A$ is the albedo of the dust grains, $f$ is the filling factor, and $\theta$ is the observation phase angle.
Under these conditions, the dust brightness follows a $\rho^{-1}$ profile, so that $A(\theta)f\rho$ is independent of the measuring aperture size.
We derive $A(\theta)f\rho$ values from V filter images, which are assumed to be mostly free of significant gas contamination and thus dominated by dust-scattered solar continuum.
Additional correction via the Halley-Marcus curve \citep{SchleicherBair2011} to a phase factor of zero is applied to yield $A(0)f\rho$.

The dust content is estimated with the magnitude derived from V filter observations in a circular aperture of radius $\rho$ via the $A(\theta)f\rho$ parameter \citep{AHearn_1984}:
\begin{equation}
    A(\theta)f\rho = 4 \cdot r_h^2 \cdot \Delta^2 \cdot 10^{0.4\left(m_{\odot} - m_V\right)} \cdot \rho^{-1}
\end{equation}
where $r_h$ is the heliocentric distance in AU, $\Delta$ is the geocentric distance, $m_{\odot}$ is the magnitude of the sun calculated by convolving a solar spectral model \citep{Colina_1996} taken from the \verb|sbpy| python package \citep{Mommert2019} with the effective area of the V filter, and $m_V$ is the magnitude of the V-band image inside the aperture.
We use the extracted brightness profile of the V-band image for a radial characterization of $A(\theta)f\rho$ when we have sufficient spatial resolution, and fall back to a pair of $1\times10^4$ km and $1\times10^5$ km apertures otherwise.

\subsection{Active Area}
The sublimation model employed is based on the formalism of \cite{CowanAHearn1979}, which assumes a rapidly rotating nucleus with sufficiently high thermal inertia such that each parallel of latitude is in thermal equilibrium.
We adopt typical cometary surface parameters, with a visual albedo of 0.05 and an infrared emissivity $\epsilon = 1.0$ to derive an average molecular flux of H$_2$O molecules leaving the surface, denoted by $Z$.
Additionally, the north pole of the comet is pointed along the comet-sun axis for maximal heating and sublimation.
The resulting active area, Q(H$_2$O)/$Z$, provides a first-order minimum estimate of the surface area of the nucleus contributing to sublimation under these assumptions.

\label{sec:results}
\section{Results}

\subsection{Aperture Photometry}
For large $r_h$, we report the details of aperture photometry applied to the stacked images of each epoch: the combination of weak signal and color uncertainty have the largest impact on the interpretation of the results for this section of the apparition.
As mentioned in Section \ref{sec:stat_approach_reddening}, in each epoch we assess the influence of the assumed prior on $S'$ by repeating the analysis over a grid of candidate priors, with $\mu_{S'}$ spaced in steps of 1\%/1000 \AA.
From this larger data set, for each epoch we report the priors that result in 1$\sigma$, 3$\sigma$, and 5$\sigma$ detections, followed by the results based on the reference prior stated in Section \ref{sec:lpc_stats}.
Inbound production of OH beyond 3 AU is detected at significance, but disappears post-perihelion unless there is significant reddening of the coma in its outbound phase.
From the range of priors reported, it is clear that our lack of information of $S'$ generally dominates our derived production values and detection confidence.
The complete aperture photometry summary is shown in Table \ref{table:aperture_photometry_summary}.

\startlongtable
\begin{deluxetable}{cccccccccc}
    \tablewidth{0pt}
    \tablecaption{Aperture photometry summary \label{table:aperture_photometry_summary}}
    \tablehead{\colhead{Epoch} & $\mu_{S'}$ & SNR & $\left<\dot{N}\right>$ & $\left<\text{OH Flux}\right>$ & $\left<\text{N(OH)}\right>$ & $\left<\text{Q(OH)}\right>$ & $\hat{r}$ & $\left<\text{Q}(\text{H}_2\text{O})\right>$ & $\left<\text{Q}_{MLE}\text{(H}_2\text{O)}\right>$ \\ & \%/1000\AA\ & & counts~s$^{-1}$ & $10^{-14}$ erg/s/cm$^2$ & $10^{32}$ & $10^{26}$ s$^{-1}$ & \% & $10^{27}$ s$^{-1}$ & $10^{25}$ s$^{-1}$}
    \startdata
    1 & 36.0 & 1.0 & 0.044 $\pm$ 0.036 & 5.67 $\pm$ 4.64 & 0.71 $\pm$ 0.58 & 0.16 $\pm$ 0.13 & 12.1 & 2.7 $\pm$ 2.2 & 1.8 $\pm$ 1.4 \\ 
    & 42.0 & 3.0 & 0.110 $\pm$ 0.036 & 14.02 $\pm$ 4.60 & 1.76 $\pm$ 0.58 & 0.40 $\pm$ 0.13 & 0.1 & 6.5 $\pm$ 2.1 & 4.4 $\pm$ 1.4 \\
    & 49.0 & 5.0 & 0.179 $\pm$ 0.036 & 22.87 $\pm$ 4.57 & 2.87 $\pm$ 0.57 & 0.65 $\pm$ 0.13 & 0.0 & 10.7 $\pm$ 2.1 & 7.1 $\pm$ 1.4 \\
    & 20.0 & -4.4 & -0.164 $\pm$ 0.038 & -20.87 $\pm$ 4.80 & -2.62 $\pm$ 0.60 & -0.60 $\pm$ 0.14 & 100.0 & -9.7 $\pm$ 2.2 & -6.5 $\pm$ 1.5 \\
    \hline 
    2 & 14.0 & 1.0 & 0.079 $\pm$ 0.051 & 10.03 $\pm$ 6.56 & 0.93 $\pm$ 0.61 & 0.28 $\pm$ 0.19 & 20.1 & 2.9 $\pm$ 2.0 & 3.1 $\pm$ 2.0 \\
    & 17.0 & 3.0 & 0.155 $\pm$ 0.051 & 19.81 $\pm$ 6.50 & 1.83 $\pm$ 0.60 & 0.56 $\pm$ 0.18 & 3.3 & 5.8 $\pm$ 1.9 & 6.1 $\pm$ 2.0 \\
    & 22.0 & 5.0 & 0.274 $\pm$ 0.050 & 34.99 $\pm$ 6.42 & 3.23 $\pm$ 0.59 & 0.99 $\pm$ 0.18 & 0.0 & 10.4 $\pm$ 1.9 & 10.8 $\pm$ 2.0 \\
    & 20.0 & 4.5 & 0.228 $\pm$ 0.051 & 29.07 $\pm$ 6.45 & 2.69 $\pm$ 0.60 & 0.82 $\pm$ 0.18 & 0.2 & 8.6 $\pm$ 1.9 & 9.0 $\pm$ 2.0 \\
    \hline
    3 & 13.0 & 1.0 & 0.085 $\pm$ 0.068 & 10.88 $\pm$ 8.65 & 0.96 $\pm$ 0.77 & 0.36 $\pm$ 0.28 & 20.1 & 2.9 $\pm$ 2.3 & 3.9 $\pm$ 3.1 \\
    & 17.0 & 3.0 & 0.220 $\pm$ 0.067 & 28.05 $\pm$ 8.55 & 2.49 $\pm$ 0.76 & 0.92 $\pm$ 0.28 & 1.5 & 7.5 $\pm$ 2.3 & 10.0 $\pm$ 3.1 \\
    & 21.0 & 5.0 & 0.345 $\pm$ 0.066 & 44.03 $\pm$ 8.47 & 3.91 $\pm$ 0.75 & 1.45 $\pm$ 0.28 & 0.0 & 11.9 $\pm$ 2.3 & 15.7 $\pm$ 3.0 \\
    & 20.0 & 4.7 & 0.315 $\pm$ 0.067 & 40.14 $\pm$ 8.49 & 3.56 $\pm$ 0.75 & 1.32 $\pm$ 0.28 & 0.1 & 10.8 $\pm$ 2.3 & 14.3 $\pm$ 3.0 \\
    \hline
    4 & 7.0 & 1.0 & 0.164 $\pm$ 0.119 & 20.89 $\pm$ 15.14 & 0.40 $\pm$ 0.29 & 0.31 $\pm$ 0.23 & 43.4 & 0.9 $\pm$ 0.7 & 3.4 $\pm$ 2.5 \\
    & 8.0 & 3.0 & 0.382 $\pm$ 0.118 & 48.69 $\pm$ 15.10 & 0.94 $\pm$ 0.29 & 0.73 $\pm$ 0.23 & 30.8 & 2.1 $\pm$ 0.7 & 8.0 $\pm$ 2.5 \\
    & 10.0 & 5.0 & 0.805 $\pm$ 0.118 & 102.69 $\pm$ 15.01 & 1.99 $\pm$ 0.29 & 1.54 $\pm$ 0.23 & 12.1 & 4.6 $\pm$ 0.7 & 16.8 $\pm$ 2.5 \\
    & 20.0 & 23.5 & 2.701 $\pm$ 0.115 & 344.38 $\pm$ 14.64 & 6.67 $\pm$ 0.28 & 5.18 $\pm$ 0.22 & 0.0 & 15.6 $\pm$ 0.7 & 56.2 $\pm$ 2.4 \\
    \hline
    11 & 25.0 & 1.0 & 0.061 $\pm$ 0.049 & 7.78 $\pm$ 6.21 & 0.13 $\pm$ 0.10 & 0.06 $\pm$ 0.05 & 20.1 & 0.3 $\pm$ 0.3 & 0.7 $\pm$ 0.5 \\
    & 29.0 & 3.0 & 0.162 $\pm$ 0.048 & 20.67 $\pm$ 6.15 & 0.34 $\pm$ 0.10 & 0.16 $\pm$ 0.05 & 1.5 & 0.9 $\pm$ 0.3 & 1.8 $\pm$ 0.5 \\
    & 33.0 & 5.0 & 0.257 $\pm$ 0.048 & 32.76 $\pm$ 6.10 & 0.54 $\pm$ 0.10 & 0.26 $\pm$ 0.05 & 0.0 & 1.5 $\pm$ 0.3 & 2.8 $\pm$ 0.5 \\
    & 20.0 & -1.5 & -0.075 $\pm$ 0.049 & -9.61 $\pm$ 6.29 & -0.16 $\pm$ 0.10 & -0.08 $\pm$ 0.05 & 79.9 & -0.4 $\pm$ 0.3 & -0.8 $\pm$ 0.5 \\
    \hline
    12 & 20.0 & 1.0 & 0.049 $\pm$ 0.046 & 6.21 $\pm$ 5.87 & 0.14 $\pm$ 0.13 & 0.04 $\pm$ 0.04 & 20.1 & 0.4 $\pm$ 0.4 & 0.5 $\pm$ 0.4 \\
    & 26.0 & 3.0 & 0.142 $\pm$ 0.045 & 18.14 $\pm$ 5.75 & 0.40 $\pm$ 0.13 & 0.12 $\pm$ 0.04 & 0.2 & 1.3 $\pm$ 0.4 & 1.3 $\pm$ 0.4 \\
    & 32.0 & 5.0 & 0.227 $\pm$ 0.044 & 28.94 $\pm$ 5.66 & 0.63 $\pm$ 0.12 & 0.19 $\pm$ 0.04 & 0.0 & 2.1 $\pm$ 0.4 & 2.1 $\pm$ 0.4 \\
    & 20.0 & 1.0 & 0.049 $\pm$ 0.046 & 6.21 $\pm$ 5.87 & 0.14 $\pm$ 0.13 & 0.04 $\pm$ 0.04 & 20.1 & 0.4 $\pm$ 0.4 & 0.5 $\pm$ 0.4 \\
    \enddata
    \tablecomments{Aperture photometry for epochs with limited or no spatial profile for fitting.  Quoted errors are 1-$\sigma$ stochastic errors for all sources except color uncertainty. $\mu_{S'}$ is the assumed peak of the reddening prior similar to Equation \ref{eq:bayes_dust_prior}.  SNR: signal-to-noise ratio.  Brackets indicate average values after marginalization over $S'$ with the given prior.  $\dot{N}$: excess count rate after continuum subtraction, as in Equation \ref{eq:count_rate_to_flux}, before conversion to flux and before rejection of unphysical count rates.  N(OH): number of OH molecules within the aperture.  Q(OH): steady-state OH production rate via dividing by scaled lifetime of OH.  $\hat{r}$: percentage of comets within the prior distribution that result in negative count rates.  Q(H$_2$O): vectorial forward-modeled water production rate by matching N(OH) within the aperture. Q$_{MLE}$(H$_2$O): vectorial forward-modeled water production rate resulting in the same mass loss of OH as observed within the aperture.}
\end{deluxetable}

\subsection{Water Production Curve}
Table \ref{table:production_rate_summary} shows the combined results of derived water production from column density profiles and aperture photometry.
All quantities are derived using a dust color prior $\mu_{S'}$ taken from Equation \ref{eq:bayes_dust_prior}.
The columns $\left<\text{Q(OH)}\right>$, $\left<\text{Q}_{ap}\text{(H}_2\text{O)}\right>$, and $\left<\text{Q}_{\text{MLE}}(\text{H}_2\text{O})\right>$ are taken from aperture photometry on the stacked images, while $\left<\text{Q(H}_2\text{O)}\right>$ and $\left<\text{Q}^*\text{(OH)}\right>$ are derived with vectorial model fits and Equation \ref{eq:blue_spot_integral} respectively on the symmetrical component of the stacked images.

\startlongtable
\begin{deluxetable}{cccccc}
    \tablewidth{0pt}
    \tablecaption{Summary of the production rates (in units of molecules s$^{-1}$)} \label{table:production_rate_summary}
    \tablehead{\colhead{Epoch} & $\left<\text{Q(H}_2\text{O)}\right>$ & $\left<\text{Q}^*\text{(OH)}\right>$ & $\left<\text{Q(OH)}\right>$ & $\left<\text{Q}_{ap}\text{(H}_2\text{O)}\right>$ & $\left<\text{Q}_{\text{MLE}}(\text{H}_2\text{O})\right>$\\ & vectorial & & $10^5$ km aperture & forward-modeled & }
    \startdata
    1 & - & - & - & - & -\\
    2 & - & - & $8.22 \pm 1.83 \times 10^{25}$ & $8.60 \pm 1.92 \times 10^{27}$ & $9 \pm 2 \times 10^{25}$ \\
    3 & - & - & $1.32 \pm 0.28 \times 10^{26}$ & $1.08 \pm 0.23 \times 10^{28}$ & $1.43 \pm 0.3 \times 10^{26}$\\
    4 & - & - & $5.18 \pm 0.22 \times 10^{26}$ & $1.56 \pm 0.07 \times 10^{28}$ & $5.62 \pm 0.24 \times 10^{26}$\\
    5 & $1.00 \pm 0.16 \times 10^{29}$ & $3.1 \times 10^{26}$ & $5.21 \pm 0.02 \times 10^{27}$ & $8.89 \pm 0.03 \times 10^{28}$ & -\\
    6 & $2.49 \pm 0.14 \times 10^{29}$ & $2.8 \times 10^{26}$ & $2.91 \pm 0.002 \times 10^{28}$ & $2.35 \pm 0.002 \times 10^{29}$ & -\\
    7 & $2.05 \pm 0.24 \times 10^{29}$ & $5.9 \times 10^{26}$ & $3.52 \pm 0.002 \times 10^{28}$ & $2.01 \pm 0.001 \times 10^{29}$ & -\\
    8 & $6.41 \pm 0.09 \times 10^{28}$ & $5.8 \times 10^{25}$ & $9.00 \pm 0.005 \times 10^{27}$ & $6.34 \pm 0.004 \times 10^{28}$ & -\\
    9 & $2.82 \pm 0.02 \times 10^{28}$ & $3.4 \times 10^{25}$ & $2.46 \pm 0.007 \times 10^{27}$ & $2.88 \pm 0.008 \times 10^{28}$ & -\\
    10 & $1.61 \pm 0.03 \times 10^{28}$ & - & $7.24 \pm 0.06 \times 10^{26}$ & $1.22 \pm 0.01 \times 10^{28}$ & -\\
    11 & - & - & - & - & -\\
    12 & - & - & $4.16 \pm 3.93 \times 10^{24}$ & $4.44 \pm 4.19 \times 10^{26}$ & $5 \pm 4 \times 10^{24}$\\
    \enddata
    \tablecomments{All bracketed values derived after marginalization over $S'$, using the fixed dust color prior given in Equation \ref{eq:bayes_dust_prior}.  Q(H$_2$O): water production rate after fitting outer-coma OH column density to a vectorial model, 1$\sigma$ fitting error.  Q$^*$(OH): near-nucleus excess signal of Q(H$_2$O) vectorial model, interpreted as 3080~\AA OH emission, converted to production rate with scaled lifetime.  Q(OH): production rate of OH inside stated aperture only, with 1$\sigma$ stochastic error.  Q$_{ap}$(H$_2$O): vectorial forward-modeled water production rate by matching N(OH) within the aperture, 1$\sigma$ stochastic error.  Q$_{MLE}$(H$_2$O): as in Table \ref{table:aperture_photometry_summary}, computed only at $r_h$ where nuclear sublimation of water is inefficient.}
\end{deluxetable}

\begin{figure*}
    \centering
    \includegraphics[width=1.0\linewidth]{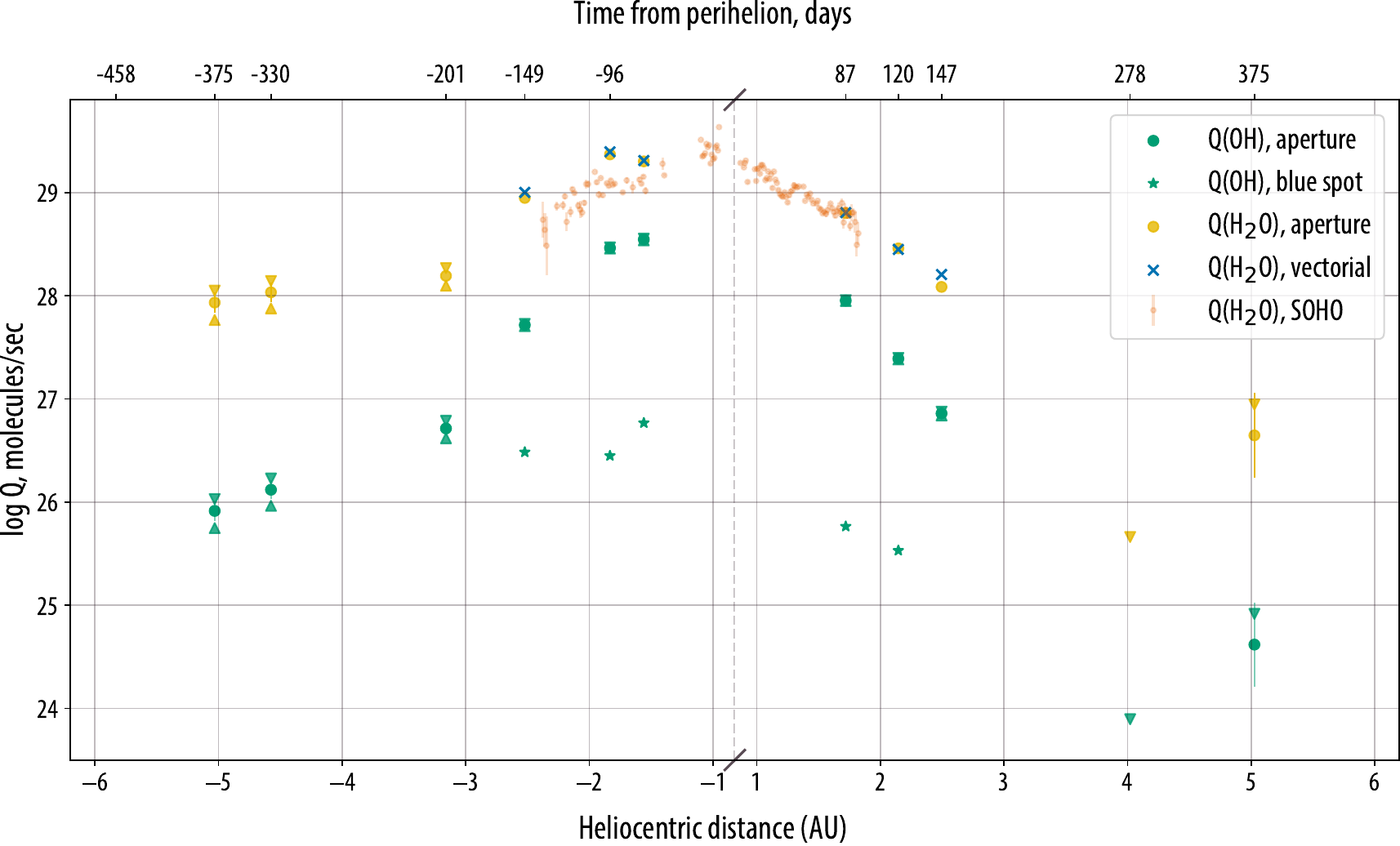}
    \caption{Water and hydroxyl production of C/2013 US$_{10}$ during its apparition.  Orange: SOHO water production rates as reported in \citet{Combi_2018}. Green: OH production measured within a $1 \times 10^5$ km aperture.  Yellow: forward-modeled water production resulting from OH measured in green.  Blue: water production derived by vectorial fits of OH column density in the outer coma.  Vertical lines indicate $1\sigma$ stochastic error, with 1$\sigma$ dust color error delimited with up/down triangles.  All \textit{Swift}-derived values shown assume a Bayesian prior for the dust color as given by Eq. \ref{eq:bayes_dust_prior}.  The Q(OH) blue spot values here interpret near-nucleus UV excess as entirely 3085 \AA~OH emission.}
    \label{fig:us10_lightcurve}
\end{figure*}

\subsection{Active Area}

\begin{figure*}
    \centering
    \includegraphics[width=1.0\linewidth]{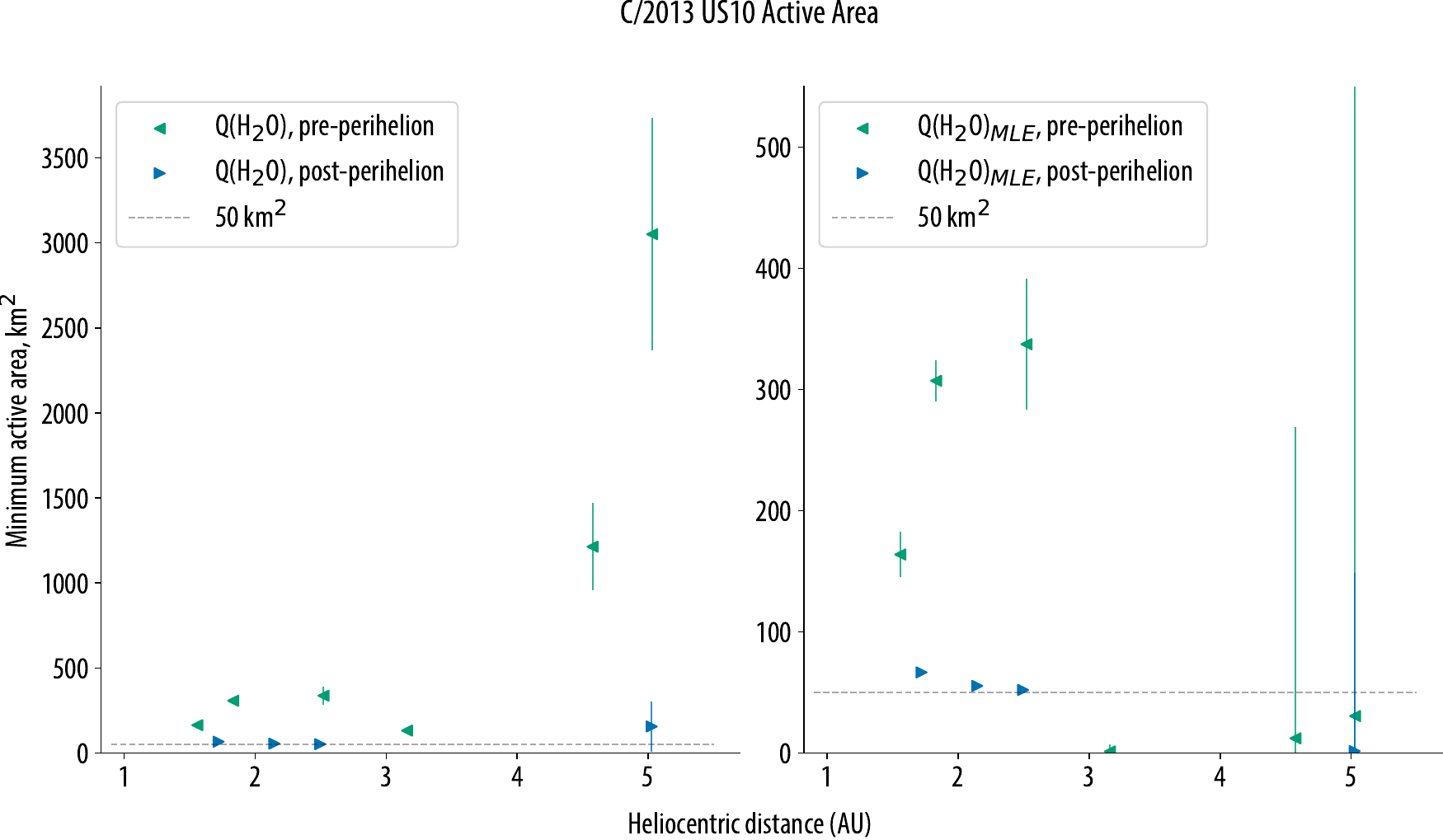}
    \caption{Minimum active area of C/2013 US$_10$, in km$^2$, over its apparition.  All values derived using the dust color prior given in Equation \ref{eq:bayes_dust_prior}.  Within 3 AU, active areas calculated with water production rates given by vectorial model fits.  Left: beyond 3 AU, active areas based on forward-modeled aperture-derived water production rates.  Right: beyond 3 AU, active areas based on mass-loss equivalent Q$_{MLE}$(H$_2$O).  The dashed horizontal line indicates an active area of 50 km$^2$.  Green and blue denote pre- and post-perihelion values respectively.}
    \label{fig:us10_active_area}
\end{figure*}

The active area computed over the apparition of C/2013 US$_{10}$ is presented in Figure \ref{fig:us10_active_area}, using water production rates derived with the dust-color prior given in Equation \ref{eq:bayes_dust_prior}. The comet exhibits clear hyperactivity on the inbound leg, with active areas substantially exceeding post-perihelion values. After perihelion, the active area settles to a near-constant plateau of $\sim$50 km$^2$.
This level of activity is consistent with strongly active long-period comets \citep{AHearn1995p27}.
A similar evolution was observed in C/2009 P1 (Garradd) \citep{Bodewits_2014a}, where \textit{Swift} measurements show pre-perihelion active areas approaching $\sim$400 km$^2$, declining to $\sim$75 km$^2$ post-perihelion.
Likewise, the Halley-type comet 12P/Pons-Brooks exhibits a stable minimum active area of $\sim$70 km$^2$ \citep{Li_2025}, indicating that the $\sim$50 km$^2$ plateau observed here is not unusual.

The primary outlier occurs on the inbound leg beyond 3 AU, where the derived column density is poorly reproduced by the vectorial model.
The resulting anomalously low active area likely reflects model limitations rather than a true decrease in activity. However, depletion of a secondary water source, such as icy grains contributing at $r_h > 4$ AU, cannot be ruled out.
If we instead consider the active area based on the water production given by Q(H$_2$O)$_{MLE}$, we have a drastically different picture: the hyperactivity begins after water sublimation becomes efficient and the surface material is being processed away.
Our active areas thus heavily depend on the modeling and assumptions we use to convert observed OH to inferred water production rates.

The post-perihelion plateau of $\sim$50 km$^2$ corresponds to an equivalent spherical surface with radius $\gtrsim$2.0 km.
Assuming a canonical active fraction of 5 to 20\%, this implies a nucleus radius of $\sim$2 to 9 km, which would place it among the larger nuclei of long period comets \citep{2017AJ15453B}.
No direct size constraints currently exist for C/2013 US$_{10}$.
A more detailed analysis could incorporate a Bayesian prior on nucleus radius to derive a statistically robust estimate of the active fraction and is left for future work.

\subsection{Dust Content}

\startlongtable
\begin{deluxetable*}{cccccccc}
\tablecolumns{8}
\tablewidth{0pt}
\tablecaption{Aperture-derived $Af\rho$ (cm) Measurements by Epoch and Projected Size\label{table:afrho_epochs}}
\tablehead{\colhead{Epoch} & \colhead{$r_{\mathrm{h}}$} & \colhead{Observation Time} & \colhead{Phase Angle} & \multicolumn{2}{c}{$\rho = 10^4$ km} & \multicolumn{2}{c}{$\rho = 10^5$ km} \\
\cline{5-6}\cline{7-8} \colhead{} & \colhead{AU} & \colhead{UTC} & \colhead{degrees} & \colhead{Af$\rho$} & \colhead{A(0)f$\rho$} & \colhead{Af$\rho$} & \colhead{A(0)f$\rho$}}
\startdata
1 & 5.83 & 2014-08-14 & 5.2 &  3791 &  4652 &  1218 &  1494 \\
2 & 5.03 & 2014-11-05 & 10.9 &  3730 &  5588 &  1368 &  2049 \\
3 & 4.57 & 2014-12-19, 2014-12-22 & 10.8 &  3876 &  5795 &  1597 &  2388 \\
4 & 3.16 & 2015-04-28 & 15.2 &  5354 &  9159 &  2223 &  3802 \\
5 & 2.52 & 2015-06-19, 2015-06-20 & 23.3 &  2903 &  6137 &  2335 &  4937 \\
6 & 1.83 & 2015-08-11, 2015-08-12 & 28.3 &  2063 &  4832 &  3744 &  8770 \\
7 & 1.56 & 2015-09-01 & 40.3 &  1591 &  4393 &  4627 & 12771 \\
8 & 1.72 & 2016-02-11 & 32.9 &  1464 &  3702 &  2157 &  5454 \\
9 & 2.15 & 2016-03-14 & 27.1 &  1754 &  4016 &  1666 &  3813 \\
10 & 2.50 & 2016-04-10 & 19.8 &  2345 &  4553 &  1592 &  3091 \\
11 & 4.02 & 2016-08-19 & 12.6 &  1179 &  1865 &   893 &  1413 \\
12 & 5.03 & 2016-11-24 & 6.6 &   863 &  1117 &   778 &  1008 \\
\enddata
\end{deluxetable*}

\begin{figure*}
    \centering
    \includegraphics[width=1.0\linewidth]{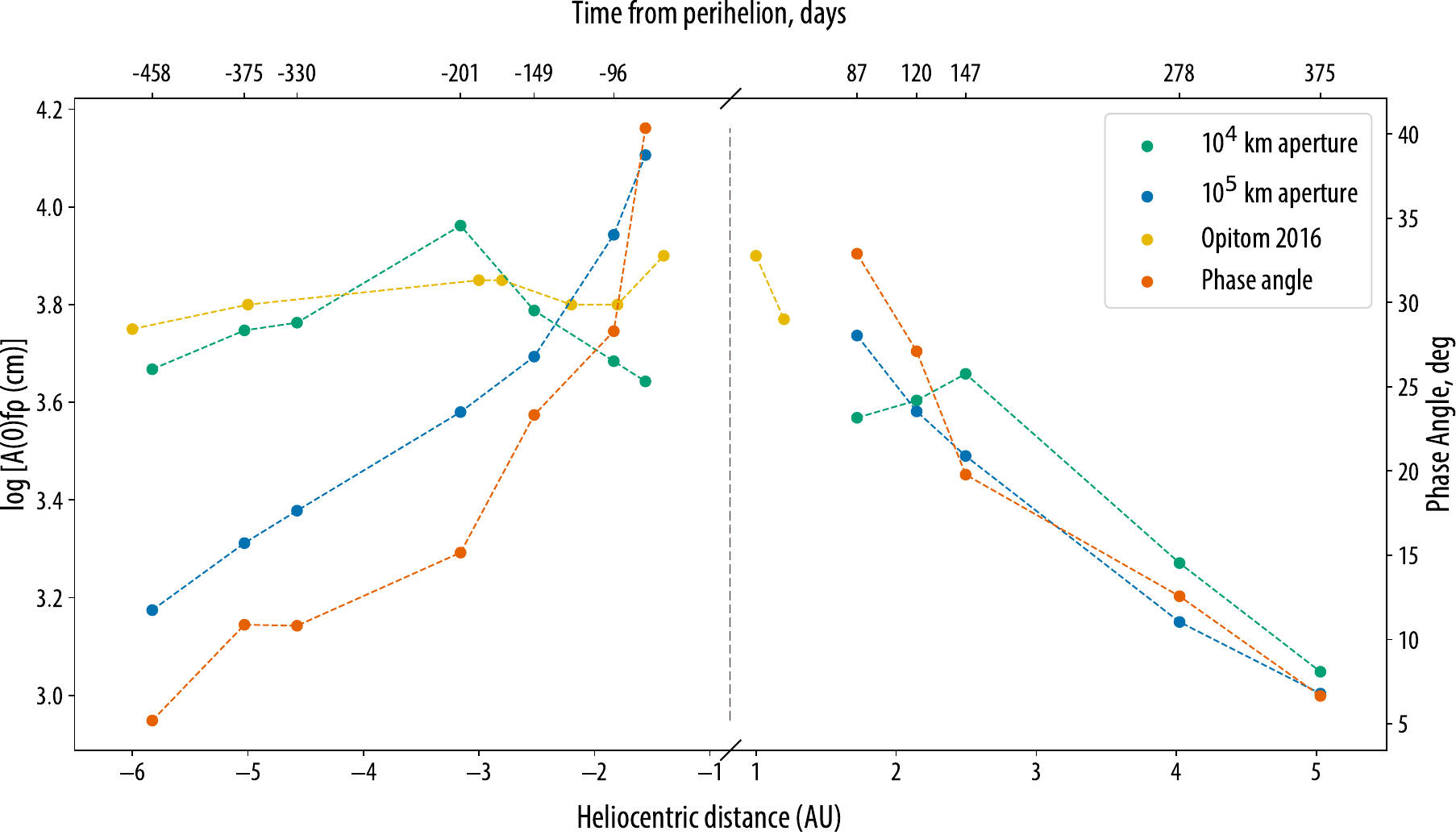}
    \centering
    \includegraphics[width=1.0\linewidth]{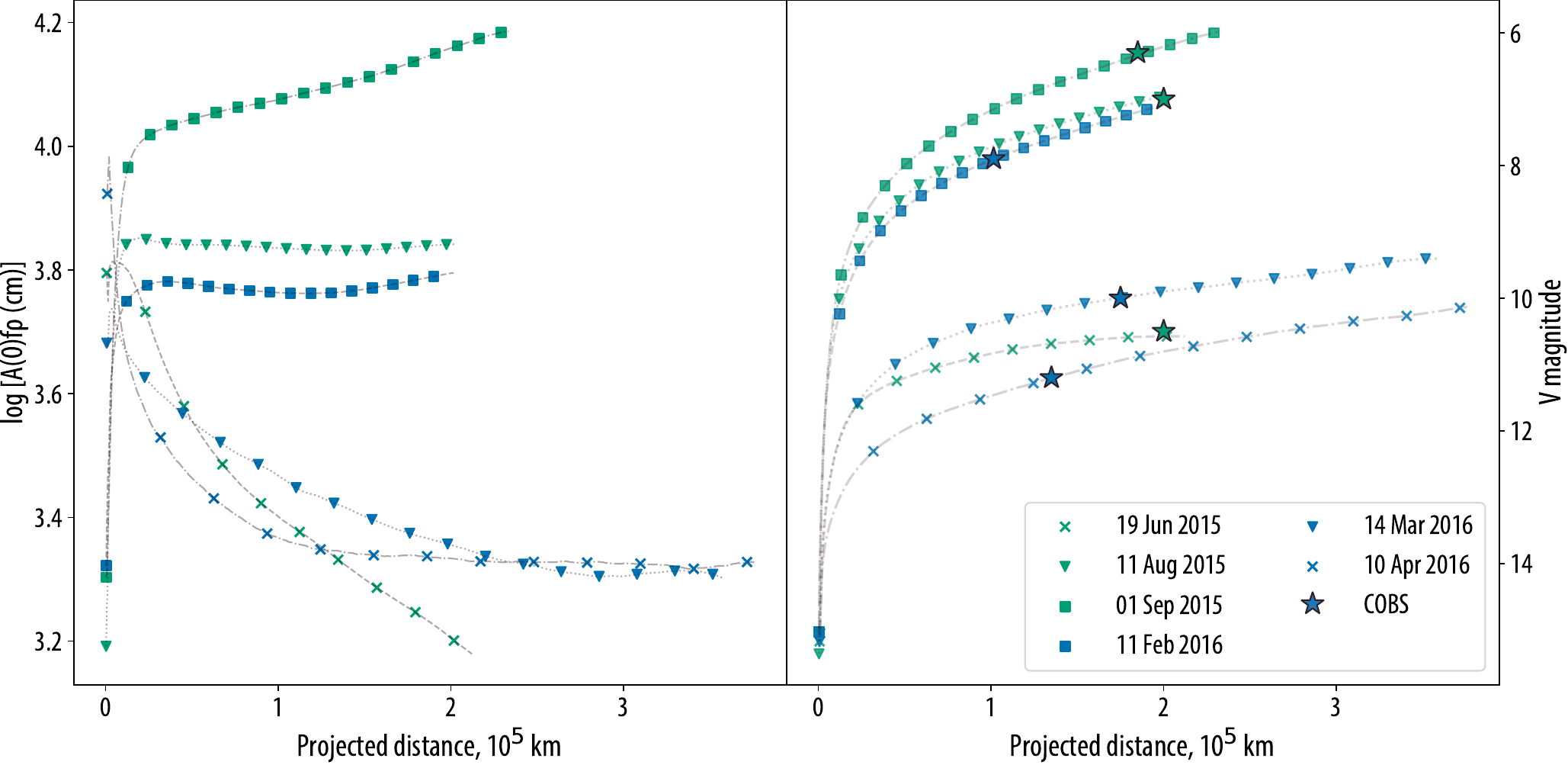}
    \caption{Top: Aperture-derived $A(0)f\rho$ values based on V magnitude during apparition for $10^4$ km and $10^5$ km projected aperture radius, along with corresponding phase angle during observation.  Opitom: $10^4$ km aperture via TRAPPIST. Bottom left: Profile-derived $A(0)f\rho$ values based on V magnitude.  Bottom right: Profile-derived V magnitude with reference magnitudes via COBS.  Radial coordinates for COBS references chosen to match their corresponding curve.  Pre- and post-perihelion profile data shown as green and blue respectively.}
    \label{fig:afrho_apertures_and_profiles}
\end{figure*}

Aperture-derived $A(0)f\rho$ values over the apparition of C/2013 US$_{10}$ are presented in Table \ref{table:afrho_epochs} and Figure \ref{fig:afrho_apertures_and_profiles}, derived from the V-band UVOT images and corrected to zero phase via the Marcus-Halley curve \citep{SchleicherBair2011}.
Results generally agree with TRAPPIST observations values found in \citet{Opitom2016}, measured in a $10^4$ km aperture.
A notable aperture-dependent effect emerges between the $10^4$ km and $10^5$ km apertures: the larger aperture shows increasing dust activity with decreasing $r_h$, while the smaller aperture sees a slight suppression within 3~AU, coinciding with the increase in efficiency of water sublimation from the nucleus.
The TRAPPIST $A(0)f\rho$ values for C/2013 US$_{10}$ also exhibit this feature, though less pronounced, and similar behavior in $A(0)f\rho$ was noted for comet C/2017 K2 in both \citet{Hmiddouch_2025} and \citet{Kwon_2024}.
It is noted in \citet{Opitom2016} that the fall in $A(0)f\rho$ within 3~AU is not unusual, and was additionally seen in C/2009 P1 (Garradd) and C/2013 A1 (Siding Spring).

For the epochs with sufficient spatial resolution, we use the radial profile pulled from the symmetric component of the V-band image to derive $A(0)f\rho$.
Closest to perihelion, there is strong indication of steady dust production from the flat $A(0)f\rho$ profile as a function of $\rho$.
The September 2015 (epoch 7) observation suffers from an inability to properly identify background, showing instead a linear increase from the background of the V-band.
We presume the true shape of epoch 7's $A(0)f\rho$ profile to be flat based on the preceding observation in August 2015 (epoch 6) and the post-perihelion follow-up in Feb 2016 (epoch 8) showing stable outflow over this leg of the apparition.
The azimuthally-averaged $A(0)f\rho$ in \citet{Woodward_2021} of C/2013 US$_{10}$ for January 11, 2016 gives $\log A(0)f\rho$ to be roughly 3.78, while the terminal value of $\log A(0)f\rho$ from epoch 8 (one month later) is in good agreement at 3.79.

Significantly before and after perihelion, however, we see a strong fall-off of $A(0)f\rho$ with increasing $\rho$.
In the post-perihelion epochs 9 and 10 (March and April 2016), log$A(0)f\rho$ (cm) settles to a common value of $\approx 3.3$ at $5\times10^5$ km and $3\times10^5$ km respectively, consistent with accelerating smaller grains in the inner coma and a stable large-grain population in the outer coma \citep{Agarwal_2024}.
In the pre-perihelion case in June 2015 (epoch 5), no such flat region for $A(0)f\rho$ occurs, indicating continual dust evolution with $\rho$.
It should be noted that epoch 5 also exhibited a column density that could not be well fit with a vectorial model, strongly suggesting extended sources and a disappearing grain population that could possibly drive the observed profile behavior.

\section{Discussion}\label{sec:discussion}
The \textit{Swift} observation campaign of C/2013 US$_{10}$ provides an opportunity to study a dynamically new Oort-cloud comet over an unusually broad portion of its apparition, from large heliocentric distances before perihelion to similarly large distances afterward.
In a broader context, such coverage is useful because activity at these distances need not evolve symmetrically and may reflect changing contributions from different volatile reservoirs, seasonal illumination, and thermal history.
At the same time, overlap with independent observations allows us to test the fidelity of the new analysis pipeline against an established observational record.
We therefore view the C/2103 US$_{10}$ data set both as a probe of the comet’s activity over a wide dynamical range and as a validation case, given that the pipeline recovers values that are consistent with established literature.

\subsection{Water Production}\label{sec:discussion_water_production}
Our results within the water ice line show reasonable agreement with Q(H$_2$O) values with those obtained via SOHO/SWAN observations of extended hydrogen Lyman-alpha emission \citep{Combi_2018}, though our values pre-perihelion are slightly larger, likely due to backgrounding issues as the comet coma occupies the entire UVOT image.
The agreement between profile fitting and forward-modeled aperture water production rates within 3 AU has two important implications.
First, it indicates that the asymmetric components of the images do not significantly affect our water production rates, as those components are captured in aperture signals but not in our extracted brightness profiles used for vectorial fitting.
Second, the forward-modeling based on the $10^5$ km apertures is not significantly affected by the excess flux near the nucleus, which is not surprising given that it is roughly two to three orders of magnitude smaller than the overall flux in the aperture.
However, this does suggest that smaller apertures may be more sensitive to the excess flux, and profile fitting that focuses on the inner coma can overestimate water production values due to the near-nucleus enhancement in column density.

At large heliocentric distances, the assumptions of the underlying vectorial model used may not hold: in particular, water in the coma is assumed to be exclusively the result of uniform outflow from the nucleus, with no significant contribution from jets, extended sources, or icy grains.
Both the vectorial and Haser models infer a distribution of water and OH outside an aperture based on measurements inside the aperture; any additional water sources, such as slow-moving evaporating icy grains, in the aperture can cause the model to infer much more water outside the aperture than is necessarily present.
Possible examples include C/2009 P1 (Garradd) and 3I/ATLAS, for which, at comparable heliocentric distances, larger-aperture water production rates measured by \textit{Swift} \citep{Bodewits_2014a,xing2025_3i} exceed narrower-slit spectroscopic estimates \citep{Paganini_2012,Hutsemekers_2026}, potentially reflecting different sensitivities to extended water sources.
Distributed sources are unaccounted for in the Haser and vectorial models, and result in aperture-dependent distortions to production rates.

We infer the total water production rate, Q(H$_2$O), by forward-modeling the signal measured within a given aperture, incorporating the commonly adopted empirical relation for the outflow velocity of water, $v = 0.85/\sqrt{r_h}\ \mathrm{km\,s^{-1}}$ \citep{Cochran_Schleicher_1993}.
The relation assumes that the bulk gas driving the activity is water, which may not be the case at large heliocentric distances \citep{2012ApJ75215O}.
Even for observations at $r_h > 3$ AU, the vectorial model predicts an extended OH coma via the OH velocity kick (Section \ref{sec:vectorial_model}), including substantial flux far outside our standardized aperture, which is inconsistent with our images (Top panel of Figure \ref{fig:showcase}).
If the observable OH signal is confined mainly to the inner coma at large $r_h$ as the images suggest, then converting the in-aperture measurement to a total Q(H$_2$O) under these assumptions will systematically overestimate the true water production rate.
In Table \ref{table:aperture_photometry_summary} we see strong disagreement between the forward-modeled Q(H$_2$O) and Q$_{MLE}$(H$_2$O).
Because Q$_{MLE}$ is derived from this small clump of near-nucleus OH only, this suggests that the OH seen at large $r_h$ is not well described by the vectorial model, and its compactness implies a much lower OH speed than expected from water outflow followed by photo-dissociation.
Similar to large-$r_h$ aperture measurements, the near-nucleus UV excess may originate in a distribution of OH that is inconsistent with the photo-dissociation of water.
Before we can interpret such signals, we must investigate the possible sources of the UV excess flux to determine if the excess is from UV contaminants or a compact distribution of near-nucleus OH.

\subsection{The Pale Blue Spot}\label{sec:pbs}
The derived OH column density $\Sigma_{OH}(S', \rho)$ described in Section \ref{sec:coma_split} is a function of the spectral gradient, so it was initially assumed that the excess flux was due to the dust being more blue near the nucleus.
If true, then our assumption of spatial uniformity of $S'$ would be incorrect, and referred to the excess as the `pale blue spot' (PBS for short in this work) in tribute to Voyager 1's photograph of Earth known as `Pale Blue Dot'.
The existence of the feature is robust against a changing color prior, but the signal excess can vary as a function of the mean dust redness $\mu_{S'}$ of our color prior, with errors estimated to be at least 10\%.
The physical scale of the PBS is approximately $r_{pbs} = 2\times10^4$ km, which is stable across the apparition until 90 days after perihelion, where the excess weakens by an order of magnitude before disappearing as seen in Table \ref{table:production_rate_summary}.
Although the PBS signal is interpreted as excess OH 3085 \AA~emission in the above table, the broadness and long red tail of the UVW1 filter force us to also consider UV contaminants as the source of excess flux.

The overall production rate of the PBS hovers around 2.5 to 3 orders of magnitude less than Q(H$_2$O) inferred from the vectorial model fits, so this feature is unlikely to affect aperture photometry on highly active comets.
The local column densities near the nucleus, however, can be enhanced by as much as a factor of two due to this UV excess before falling in line with vectorial model predictions at larger nucleocentric distances.
The signal of the PBS forms a consistent, continuous picture with the large-$r_h$ aperture measurements of Q(OH) inbound.
We take this as weak evidence of a single, but secondary, source of UV emission rather than a set of evolving contaminants that would inject discontinuity into the production curve.
Until the exact nature of the PBS is clear, we approach the discussion on contaminants with the assumption that the PBS and large-$r_h$ aperture signals come from a common cause, though there is a distinct possibility that the UV excess is a cumulative signal from various UV contaminants in the chemical zoo of the coma.

Using a crude slow-moving water parent outflow vectorial model to simulate icy grains cannot reproduce the spatial profile of the PBS because the vectorial kick imparted to the OH fragments upon photodissociation spreads them out over hundreds of thousands of kilometers.
The Haser model does not consider this photochemical process, and is merely a curve fit with parameters that are loosely associated with physical properties, so deviations from the model may be overlooked or hand-waved away.
However, while the vectorial model still involves a fair amount of assumptions, its parameters are observed physical properties like lifetimes and velocities.
Deviations from the vectorial model may suggest real physics or chemistry being unaccounted for, such as a secondary parent or alternative volatile release mechanism.

\subsubsection{Processing and Instrument Artifacts}
As mentioned in Section \ref{sec:data_prep}, a flux-preserving 5" radius kernel is applied to the UVOT images before stacking to account for coincidence loss.
This is a similar scale to the PSF of the UVOT detector, so a smeared near-nucleus surface brightness may be suggested as the source of the PBS as a result of the these two effects.
After perihelion, however, the PBS weakens substantially and vectorial model fits are good even near the nucleus (Figure \ref{fig:oh_excess_and_aperture_photometry}), when pre-perihelion observations at similar $r_h$ show significant UV excess.
Further, the radial scale of the PBS extends well beyond the PSF of the instrument over the comet's apparition.
Considered alongside our agreement with Q(H$_2$O), $A(0)f\rho$, and magnitude with independent data sources, we conclude that the PBS is not an instrument issue or pipeline artifact.

The close approach in 2018/2019 of 46P to Earth ($\Delta < 0.1$ AU, 1 UVOT pixel $\approx$ 50 km) presented a unique opportunity for \textit{Swift}/UVOT to observe an active comet with high spatial resolution, making this particularly valuable for both pipeline validation and investigation into the nature of the PBS.
Our initial results from these observations of 46P show a similar pattern: a small pre-perihelion UV excess inside $10^4$ km that disappears into very good vectorial fits post-perihelion.

\subsubsection{Spatial Dust Color Variation}
Our continuum subtraction method assumes a uniform color throughout the coma.
If the dust near the nucleus was more blue than the outer coma, then this would result in an under-subtraction in the areas of the bluer dust, enhancing the derived OH flux near the nucleus and explaining the deviation from vectorial model predictions.
To test if this could be source of the PBS, we allowed $S' = S'(\rho_i)$, where $\rho_i$ is the i$^{\text{th}}$ pixel of the derived column density, and allowed $S'(\rho_i)$ to vary linearly from 40\%/1000~\AA\ in the outer coma, down to as little as $S'$ = -20\%/1000~\AA\ near the nucleus.
In principle one could force any observed column density to match vectorial model predictions via careful choice of $S'(\rho_i)$.
The unknown dust color $S'$ expands into $i$ unknowns, leaving the derived column density after continuum subtraction free to be arbitrarily chosen.
Despite the freedom, our $S'(\rho_i)$ struggled to bring the observed column density in line with a vectorial model shape, and would require severe, implausible changes in $S'(\rho_i)$ when successful.
Dust color variations in other comets such as C/2017 K2 show changes of 4.5\% to 7\%/1000~\AA\ between 10,000 and 50,000 km \citep{Zhang_2022} and a slight reddening of 10\% to 15\% for C/1995 O1 (Hale-Bopp) between the nucleus and 200,000 km in the 4100~\AA\ to 5400~\AA\ wavelength range \citep{Weiler_2003}.
When the PBS is strongest, the flat A(0)$f\rho$ profiles indicate a steady dust outflow, indicating some level of uniformity and that large color changes are improbable.
We therefore conclude that dust color variation results in a minor contribution to the PBS signal.

\subsubsection{Prompt Emission and Electron Impact Excitation}\label{sec:prompt_emission}
The photo-dissociation of water is dominated by the neutral H$_2$O $\to$ OH + H channel.
An alternate channel, known as prompt emission (PE), has excited OH* produced with a branching ratio of approximately 4.0\% and a lifetime of $10^{-6}$ s \citep{Becker_1973}, closely tracing the parent water population.
The relevant OH* prompt emission bands extend between 3060 to 3270 \AA, with similar filter response to the 3080 \AA\ OH A-X band.
The direct excitation of OH traces the distribution of H$_2$O, but only exceeds fluorescence emission within  1000 km or less \citep{1986AA160L7B}, which is generally smaller than a UVOT pixel.
Prompt emission from OH* is therefore limited to the confines of the PSF, and should persist after perihelion.
Electron impact excitation of H$_2$O in the 3050 - 3300 \AA\ region is morphologically similar to PE as it also strongly traces the water population, and the associated emission band is not expected to disappear after perihelion.
Our UV excess cannot be explained by either of these mechanisms: they are too compact and are not expected to exhibit any asymmetry about perihelion.

\subsection{Possible UV Contaminants}
The asymmetry of the PBS around perihelion seems to suggest that purely physics-driven processes, such as prompt emission, may be ruled out as the cause of our excess flux.
We might find a consistent interpretation in the release and exhaustion of a reservoir of UV-emitting species that contaminate the UVW1 filter, which has a broad transmission profile and significant red tail \citep[e.g.,][]{Bodewits2023}.
Such species would need to be slow-moving or short-lived to contain the excess near the nucleus.
The PBS is derived from the symmetric component of the image, which, paired with the expected tail-like distribution of ions due to solar-wind driven acceleration, make ions less likely as the source of the excess.

\subsubsection{CO and CO$_2$ Ions}
Both CO and CO$_2$ are the largest gaseous components of cometary comae after water \citep{Biver_2024}, and neither contribute significant UV signal in the neutral state.
However, the ionized product CO$_2^+$ has the (B$^2\Sigma_u^+$ - X$^2\Pi_g$) doublet between 2884-2896~\AA, while the CO$^+$ cometary tail system (A$^2\Pi$ - X$^2\Sigma$) has emission scattered primarily along the red tail of the UVW1 filter.
In a survey of 25 comets, \citet{HarringtonPinto_2022} found that $Q_{CO_2}/Q_{H_2O} \approx 13$\%, while the ratio $Q_{CO}/Q_{H_2O}$ is even smaller at about 3\%.
Along with an earlier survey of 18 comets by \citet{2012ApJ75215O}, we have roughly three activity regimes seen in the trends of the surveys: water dominant inside 2.5 AU, CO-dominant outside 3.5 AU, with a transitional CO$_2$-heavy bridge between the two.
This behavior is inconsistent with the seemingly smooth evolution of the PBS activity, and the outbound leg of C/2013 US$_{10}$ where CO should re-emerge as dominant shows no UV activity.
This survey-level picture ignores that particular comets can show complex evolution and strong asymmetries around perihelion for both CO and CO$_2$: Q(CO) in C/2009 P1 (Garradd) increased monotonically over the apparition \citep{Bodewits_2014a}, while 2I/Borisov showed more CO than water and may have increased Q(CO) after perihelion \citep{Bodewits_2020}, and 67P showed complicated evolution of CO, CO$_2$, and H$_2$O through surface heterogeneity and seasonal effects \citep{Combi_2020}.
With this in mind, we cannot use the apparitional asymmetry of the PBS to rule out CO or CO$_2$ ions contributing to the UVW1 signal.

We can roughly estimate the flux from CO$^+$ relative to OH with the ratio of populations and fluorescence efficiencies:
\begin{equation}\label{eq:relative_flux_estimate}
    \frac{F_{CO+}}{F_{OH}} = \Gamma \frac{N_{CO+}}{N_{OH}} \frac{g_{CO+}}{g_{OH}}
\end{equation}
where $\Gamma \approx 0.12$ is the ratio of the UVW1 filter's effective area at the strongest emission feature of CO$^+$ near 4000~\AA\ relative to the effective area around the 3080~\AA\ OH band.
We use $g_{OH} = 1.49 \times 10^{-15}$ erg/s/molecule ($2.3 \times 10^{-4}$ photons/s/molecule) \citep{Schleicher_AHearn_1988}, and take $g_{CO+} \approx 1 \times 10^{-14}$ erg/s/ion \citep{Bromley_2024} for the comet tail (A$^2\Pi$ - X$^2\Sigma)$ system, both at 1 AU and ignoring the velocity-dependent Swings effect.

Using the values in the aforementioned surveys, we compute the flux in a large-$r_h$ ($\approx 6$ AU) outflow regime with Q(CO)/Q(H$_2$O) = 5 \citep{2012ApJ75215O} and a small-$r_h$ ($\approx 2$ AU) corresponding ratio of Q(CO)/Q(H$_2$O) = 0.03 \citep{HarringtonPinto_2022}.
Estimating Q(CO$^+$) from Q(CO) requires an effective ionization fraction, but this fraction is not a fixed compositional parameter.
CO$^+$ can be produced by photoionization, solar-wind impact ionization, charge exchange, and possibly pathways involving other carbon-bearing parents, while the resulting ions are rapidly accelerated and reshaped by the solar wind.
Observations of 29P at 6 AU \citep{Cochran_1999} illustrate the difficulty: CO$^+$ production appears to be strongly influenced by solar wind charge exchange at large \(r_h\), so the effective \(Q(\mathrm{CO^+})/Q(\mathrm{CO})\) ratio depends on heliocentric distance, solar-wind conditions, aperture, and ion dynamics rather than on the neutral CO abundance alone.
Further work on 29P \citep{Ivanova_2019} suggests the main factor of the CO ionization rate is impact ionization by solar wind particles, implying a relatively low CO$^+$ production rate for C/2013 US$_{10}$ due to the low solar activity during its apparition.

We adopt the notation $f_{CO}(r_h)$ as the ionization factor and assume that CO is the only source of CO$^+$.
We approximate the population number $N$ with a small-aperture limit (relative to the scale lengths) of the Haser model so that $N \propto Q/v$, where $N$ is the number of molecules or ions in the aperture and $v$ is their velocity.
We adopt water outflow speeds via the empirical relation $v = 0.85/\sqrt r_h$ km/s and estimate the bulk ion velocity to be $v_i = 10$ km/s at large $r_h$ and $v_i = 30$ km/s \citep{Schultz_1993,Bodewits_2011} in the small $r_h$ case.
Using an order-of-magnitude estimate of Q(OH) $\sim$ Q(H$_2$O) through the branching ratio of H$_2$O + $\gamma \to$ OH + H \citep{Huebner2015ko}, our flux ratio becomes
\begin{equation}
    \frac{F_{CO+}}{F_{OH}} = \Gamma \frac{N_{CO+}}{N_{OH}} \frac{g_{CO+}}{g_{OH}} \approx f_{CO}(r_h) \cdot \Gamma \frac{Q(CO)}{Q(H_2O)} \frac{v_{H_2O}}{v_{CO+}} \frac{g_{CO+}}{g_{OH}}
\end{equation}
In the large-$r_h$ regime, this results in a flux ratio of $0.14 \cdot f_{CO}(r_h)$, while the small-$r_h$ regime yields $8.4 \times 10^{-4} \cdot f_{CO}(r_h)$.
We repeat the above calculations for the flux ratio of CO$_2^+$ with $g_{CO_2^+} = 3.89 \times 10^{-15}$ erg/s/ion ($5.6 \times 10^{-4}$ photons/sec/ion) \citep{Kim_2014} at 1 AU (no Swings effect) and Q(CO$_2$)/Q(H$_2$O) = 0.13 to get $F_{CO_2^+}/F_{OH} \approx 0.02 \cdot f_{CO_2}(r_h)$ with $\Gamma \approx 2.8$, approximated with the previously stated doublet wavelength of CO$_2$.

Reasonable ionization fractions put CO$^+$ and CO$_2^+$ at percent to sub-percent levels compared to expected OH emission.
Results from \textit{Swift}'s UVOT grism in observations of C/2007 N3 (Lulin) \citep{Bodewits_2011} showed a CO$_2^+$ flux contribution of 2\%, confirming percent-level contamination, though this was not derived with a small near-nucleus aperture but rather the much larger ion tail within the field of view.

The survey-informed changes in CO and CO$_2$ relative to water as a function of $r_h$ also pose a qualitative challenge for these ions to be our candidate for the PBS as well: the ionization fractions $f(r_h)$ must vary with $r_h$ in such a way to counteract changing CO or CO$_2$ populations to produce the suggestive continuity between aperture-derived Q(OH) and the PBS over the apparition.
While the CO$^+$ and CO$_2^+$ ions do contaminate the UVW1 filter, the low levels suggest that they are not responsible for the UV excess of the PBS.

\subsubsection{CO Cameron Bands}
An additional pathway for UV contamination by CO$_2$ is the prompt CO (a$^3\Pi$ - X$^1\Sigma$) Cameron band emission stimulated mainly via CO$_2$ photodissociation.
The modeling of \citet{Weaver_1994} suggests that this mechanism is responsible for 60\% of CO Cameron band emission, and the short lifetime of the CO a$^3\Pi$ band at $\tau \approx 7\times10^{-3}$ seconds marks this emission as a spatial tracer for the parent CO$_2$ population.
The band system stretches from 1800~\AA~ to 2600~\AA, with the largest contributions in the 2000 - 2200~\AA~region, within the bounds of the UVW1 filter.

A spectral model of CO$_2$ photodissociation (Xing 2026, in prep) yields a value of $\alpha = 9.12 \times 10^{-12}$ erg/s/count rate for the UVW1 filter attributable to CO Cameron band emission, which gives a relative flux response of $\Gamma = 7.15$ when compared to the result in Equation \ref{eq:count_rate_to_flux} for OH.
The computed g-factor for the CO Cameron bands per CO$_2$ photodissociation is $g_{CO} = 2.13 \times 10^{-6}$ photons/s/molecule ($2.02 \times 10^{-17}$ erg/s/molecule).
We then estimate the relative flux contribution to the UVW1 filter with
\begin{equation}
    \frac{F_{CO}}{F_{OH}} = \Gamma \frac{N_{CO_2}}{N_{OH}} \frac{g_{CO}}{g_{OH}} \approx \Gamma \frac{Q{(CO_2)}}{Q(H_2O)} \frac{v_{H_2O}}{v_{CO_2}} \frac{g_{CO}}{g_{OH}}
\end{equation}
where $v_{CO_2} = 0.54$ km/s when thermally scaling the water outflow relation $v = 0.85 \cdot r_h^{-\frac{1}{2}}$ to the molecular mass of CO$_2$.

The photodissociative lifetime of CO$_2$ ranges from $2 \times 10^5$ to $5 \times 10^5$ seconds during active and quiet sun conditions respectively.
At 1 AU during active sun conditions, a Haser model with lengthscale derived from the aforementioned velocity and lifetimes puts roughly 25\% of the Cameron band prompt emission column density within 20,000 km of the nucleus.
Given these conditions, the resulting flux ratio is $2.5 \times 10^{-3}$, roughly in line with the relative flux of the PBS, which is between 1 and 3 $\times 10^{-3}$ relative to OH accounted for by the vectorial model.
At larger heliocentric distances, the increase in lifetime leaves the CO Cameron bands unable to account for the excess flux: at 3 AU, just 3\% of the CO Cameron bands flux falls within the PBS.
If we consider the relatively quiet solar conditions of 2016, the PBS contribution is further reduced to roughly 10\% and 1\% of the Cameron band emission respectively, resulting in flux ratios of $1 \times 10^{-3}$ and $1 \times 10^{-4}$ at 1 AU and 3 AU.
We consider the prompt emission of CO Cameron bands a possible significant near-nucleus contaminant of the UVW1 filter at small heliocentric distances, but it cannot account for the PBS at distances outside 2 AU.

\subsubsection{Ni and Fe}
The survey of \citet{Manfroid_2021} showed that neutral Fe~I and Ni~I are ubiquitous in cometary atmospheres, including coma observations at heliocentric distances greater than 3 AU, with typical abundances relative to water on the order of $3 \times 10^{-6}$.
This result demonstrates that atomic metals can be present even in cold comae far from the Sun and therefore cannot be excluded from consideration purely on thermodynamic grounds.
Both Ni and Fe are inferred to be the product of short-lived parent molecules that have lifetimes of hundreds of seconds, keeping their emission within the spatial extent of the PBS and therefore a possible contributor to the excess UV signal.
The uncertain parentage of both species leaves pre/post-perihelion chemical evolution a distinct possibility.
The strongest reported cometary Fe~I and Ni~I lines occur mainly in the near-UV/blue optical region around 3400--3900~\AA, placing the emission firmly in the red tail of the UVW1 filter.
Fe~I and Ni~I are therefore best regarded as possible but likely secondary contaminants of UVW1, contributing primarily through the long red tail of the UVW1 filter with limited sensitivity.

We again estimate the relative contribution of photons given the relative abundances, ignoring the Swings effect, via the fluorescence efficiencies of OH and the strong lines of Ni~I and Fe~I.
We use $g_{Ni} = 6 \times 10^{-3}$ photons/s/atom at 3414.8 \AA~\citep{Bromley_2021, Bromley_2026}, and $g_{Fe} = 2.5 \times 10^{-3}$ photons/s/atom \citep{Bromley_2021, Bromley_2026}.
The relative flux is then estimated with Equation \ref{eq:relative_flux_estimate}:
\begin{equation}
    \frac{F_{Ni,Fe}}{F_{OH}} = \Gamma \frac{N_{Ni,Fe}}{N_{OH}} \frac{g_{Ni,Fe}}{g_{OH}}
\end{equation}
where $\Gamma \approx 0.2$ for both species.
For the strongest emission lines this gives a flux for Ni~I and Fe~I on the order of $1 \times 10^{-5}$ and $1 \times 10^{-6}$ relative to OH.
Even with generous summing of the Ni and Fe line forests, the total relative flux contribution reaches $\approx 10^{-4}$ and $10^{-5}$ for Ni~I and Fe~I respectively.
The fluorescence efficiency can vary by as much as a factor of 3 (in the range $-100 \text{ km/s} < \dot{r_h} < 100 \text{ km/s}$) for Ni and Fe when the Swings effect is taken into account, which is not enough of an increase to make the metals a significant contributor.

If we instead use the ratio of the excess flux of the PBS to the overall OH flux being $\approx 10^{-3}$ (Table \ref{table:production_rate_summary}), this would demand Ni~I levels an order of magnitude above what is typically seen in relation to water \citep{Manfroid_2021,Bromley_2021}.
The PBS then cannot be explained by these metallic contaminants, even if they are active in the coma at large heliocentric distances and constitute a minor contamination of UV flux in the UVW1 filter.

\subsubsection{CS}
CS is another natural contaminant to consider in the interpretation of excess UV emission in the UVW1 filter.
The CS$(0,0)$ band at 2576\,\AA\ lies near the peak of the UVW1 effective area, making the filter roughly 2--3 times more sensitive to CS than to OH.
In addition, \citet{Bromley_2025} give a fluorescence efficiency for CS$(0,0)$ that is 2.2 times larger than that of OH$(0,0)$.
Taken together, these factors imply that a given UVW1 excess can be produced by substantially fewer CS molecules than OH molecules.
Since typical cometary production ratios are of order CS$_2$/H$_2$O $\sim 2.0\times10^{-4}$ and CS/H$_2$O $\sim 3 - 20 \times 10^{-4}$ \citep{Biver_2024_comets_iii}, the inferred abundance required to explain the PBS as CS is not obviously unreasonable, and asymmetry through chemical evolution is plausible.

The present difficulty is not whether CS can be bright enough, but whether it can remain sufficiently localized near the nucleus. 
\citet{Feldman_2010} adopt a very short parent lifetime for CS$_2$ of 340~s, so fragment CS should be produced close to the nucleus, with the study using a fragment kick speed of 0.9 km s$^{-1}$. This is consistent with the average excess energy of 0.5-1.5 eV from photodissociation of CS2 into CS \citep{Huebner2015ko}.
The same work adopts a CS lifetime of $10^{5}$~s and a fluorescence efficiency for the CS$(0,0)$ band of $g_{CS} = 5.4\times10^{-4}$ photons s$^{-1}$ molecule$^{-1}$.
\citet{Biver_2011} derive a CS photodissociation rate of $2.5\times10^{-5}$ s$^{-1}$, corresponding to a lifetime of roughly $4\times10^{4}$~s, somewhat shorter than the value adopted by \citet{Feldman_2010}.
For a characteristic PBS scale of $\sim2\times10^{4}$ km, a kick speed of 0.9 km s$^{-1}$ implies a crossing time of only $\sim0.3$ d, roughly half of the shorter CS lifetime.
Using either CS lifetime, a significant fraction of the CS emission occurs outside of the PBS.

The CS radical remains a plausible but poorly constrained contaminant \citep{Noonan_2023,Cordiner_2023}.
Its band placement and fluorescence efficiency make it unusually effective at contributing to UVW1, but the spatial distribution of the fragment emission depends sensitively on fragment kinematics that are not yet well established.
If the high fragment kick speeds are correct, then CS should be too spatially extended to explain a tightly concentrated near-nucleus excess on its own.
If the kick speeds are sufficiently low, then CS could remain localized and become more difficult to rule out.
ALMA observations of 46P/Wirtanen \citet{Cordiner_2023} concluded an uncertain parentage of CS as CS$_2$ alone could not explain their derived length scales, and additionally noted that the lower velocity of CS (along with H$_2$CO and HNC) may point to slow-moving, grain-based production.
Fluorescence efficiencies that address the Swings effect in CS were only recently calculated by \citet{Bromley_2025}, though CS overall remains under-studied.
We therefore regard CS as an important systematic uncertainty in the interpretation of the UVW1 excess: CS is positioned to be a strong UV contaminant, but at highly uncertain levels.

\subsubsection{CN}\label{sec:contamination_cn}

The survey of \citet{Ochner_2025} shows a definite presence of the B-X/Violet CN emission band around 3880 \AA~ for C/2013 US$_{10}$ but the lower wavelength limit of the spectrum does not reach the 3080 \AA~ emission band of OH, so we cannot estimate the relative flux contribution from the spectrum alone.
Assuming optically thin emission, we can again use Equation \ref{eq:relative_flux_estimate} to estimate the relative flux from the CN violet line at 3880 \AA:
\begin{equation}
    \frac{F_{CN}}{F_{OH}} = \Gamma \frac{N_{CN}}{N_{OH}} \frac{g_{CN}}{g_{OH}}
\end{equation}
where $\Gamma \approx 0.13$ is the relative effective area of the UVW1 filter at 3880 \AA\ compared to 3080 \AA.
For a typical comet, published abundance ratios imply $\log(\mathrm{CN}/\mathrm{OH}) \approx -2.5$, or $N_{CN}/N_{OH} \approx 3 \times 10^{-3}$ \citep{AHearn1995p27,Hyland_2019}.
The CN fluorescence efficiency is strongly dependent both on heliocentric distance and velocity, so we approximate $\frac{g_{CN}}{g_{OH}} \approx 175$ using values at 1 AU and zero velocity.
The relative flux contribution $F_{CN}/F_{OH}$ is then 0.07, the largest contributor to flux in the UVW1 filter after OH and possibly CS.

We must then consider the spatial distribution of the CN radical and whether a significant portion of this contamination will be near the nucleus.
\citet{Fray_2005} finds that HCN to CN photodissociation models result in Haser scale lengths of roughly $1.9 \times 10^4 \cdot r_h^{1.9}$ km and $3.6 \times 10^5 \cdot r_h^2$ respectively, and that HCN alone often cannot explain the observed levels of CN.
The review of extended sources of \citet{Cottin_2008} concludes that nearly all observed CN radicals can be explained through photodissociation of HCN beyond 3 AU, and that a secondary source for CN becomes active within 3 AU.
Rosetta's in situ measurements of CN in the inner coma of 67P \citep{Hanni_2020} point to a distributed CN source, as well as jets of CN in 1P/Halley \citep{AHearn_1986} implying dust-grain based CN production.

The CN produced through photodissociation is spread over a wide volume of the coma, limiting its flux contribution near the nucleus.
If we attribute the PBS flux to CN not produced by photodissociation, we could scale the 7\% relative flux contribution of CN down proportionally, but the fraction of CN not explainable by photodissociation is currently poorly constrained.
Additionally, the presence of distributed CN sources themselves are inconsistent: only 7 out of a sample of 21 short and long period comets exhibit such a property \citep{DelloRusso_2016}.

The CN column density of C/2013 US$_{10}$ at 2.87 AU inbound in \citet{Opitom2016} suggests that CN is well-described by standard Haser modeling and does not exhibit any near-nucleus enhancements down to $10^3$ km.
This TRAPPIST measurement nearly coincides with a maximum in our PBS signal.
While this does not remove CN from consideration as a contaminant, it does heavily imply that CN is not the source of the excess UV signal in our own column densities - rather, the 7\% relative flux contamination is distributed throughout the coma in low levels.

\subsubsection{Icy Grains and Slow OH}
Icy grains are often invoked as an extended source of water to explain hyperactivity and the presence of H$_2$O in the coma beyond the water ice line \citep{AHearn_1984, 2011Sci3321396A, Bodewits_2014a, Lis2019ve}.
If we presume that the PBS is driven not by contamination but is purely excess OH (A-X) emission, then these radicals must necessarily be moving slower than OH produced by photodissociation of water to match the compact surface brightness profiles.
Emission from OH near the nucleus in the UVW1 waveband could have multiple sources.
One possibility is prompt excitation and emission of OH as discussed in Section \ref{sec:prompt_emission}, or fluorescence from slow-moving OH radicals whose photodissociative kick is somehow missing.
The slow radicals would have an uncertain parentage:
\begin{itemize}
    \item Irregularly-shaped icy dust grains, containing alcoves in which water photodissociates with the resulting OH collisionally quenched through impact with the interior grain surface,
    \item Surface sputtering, either directly from the nucleus or from grains,
    \item Complex organics or CHON breakdown,
    \item Trapped gases that directly dissociate into OH such as H$_2$O$_2$,
    \item Direct release of trapped OH from the nucleus or dissolving grains.
\end{itemize}
Measurements from the Tianma Radio Telescope (TMRT) of 3I/ATLAS obtain an OH speed of $v = 0.42 \pm 0.04$ km/s at $r_h = 2.27$ AU and $v = 0.8 \pm 0.08$ km/s at $r_h = 1.96$, both inbound \citep{Li_2026}.
The fragment OH speed should be only mildly dependent on the bulk outflow speed because of the large imparted vectorial kick of 1.05 km/s, but the OH speed of $0.42$ km/s is consistent with the CO$_2$-dominated outflow of 3I/ATLAS.
This suggests that the OH is largely not the product of photodissociation, and is instead being entrained in the bulk outflow: either after direct release from the nucleus, or desorption from the entrained icy grains.
Shortly after, the average OH velocity is nearly double as nuclear water sublimation begins to dominate and, presumably, the slow OH production is outpaced by standard H$_2$O $\to$ OH + H photodissociation.

It is plausible that our UV excess is in fact OH (A-X) emission.
The C/2013US$_{10}$ dataset includes several long exposures in the U filter centered around 3500 \AA, and two long exposures using the UVW2 filter whose peak lies at 2000 \AA.
With no further constraining information on the dust color, we transfer the assumed prior in Equation \ref{eq:bayes_dust_prior} to the U/V and UVW2/V filter sets.
After calculating each filter's response to OH emission and applying continuum removal, the derived water production values are typically within 10\% for the U filter.
The UVW2-derived water production is lower than that derived through UVW1, but still consistent within error.
The near-nucleus excess signal is sensitive to the dust redness so the scatter of values is roughly 50\%, with no clear pattern across the filter set over the apparition.
Across the UVW2/UVW1/U filter set, the derived water production values are consistent when interpreted as OH 3085 \AA~ band fluorescence.

The detections of OH (A-X) presented in Fig.~\ref{fig:us10_lightcurve} at large $r_h$ are not unprecedented: \citet{Festou_1990} reports OH at 4.9 AU during an outburst of Comet 1P/Halley, Comet Hale-Bopp (C/1995 O1) showed OH activity at 5.1 AU \citep{Schleicher_1997}, and detections in Comet Bowell (C/1980 E1) extend to 5 AU \citep{AHearn_1984}.
These were highly aniticipated and very bright comets, warranting long and deep campaigns.
The lack of numerous reports of distant OH therefore might be an observational bias.

\citet{Protopapa_2018} detected water-ice grains in C/2013 US$_{10}$ at 5.8 and 5.0 au inbound and again at 3.9 au outbound, while the coma appeared dust-dominated by 2.3 au outbound, with no evidence of water ice in the inner 1500 km of the coma.
These observations show that water-bearing grains were present in the coma at large heliocentric distance, but that their observability evolved substantially over the apparition.
Concluding that the presence of icy grains outbound decouples them from the PBS is dangerous, however: analysis of the farthest outbound images (Epochs 11 and 12) is challenging due to a crowded starfield that the inbound images lacked.
The exposure time and image counts of the last two epochs were increased accordingly to combat this pollution, but even manual re-analysis with median-stacked images of Epochs 11 and 12 give $\left<\text{Q(OH)}\right> \sim 1 \times 10^{25}$ molecules/s at a 1$\sigma$ and 2$\sigma$ SNR respectively with color prior from Equation \ref{eq:bayes_dust_prior}.
We are left with two divergent conclusions based on whether we admit these UVOT detections at large-$r_h$ outbound.
Taking these detections as legitimate points to an overall coupling between the presence of icy grains and the PBS.
If we instead dismiss Epoch 11 and 12 as statistically inadequate, we have an intriguing evolution of the PBS and its corresponding source, suggestively parallel to the split in pre/post-perihelion behavior of the active area.

Despite the ambiguity in the outbound aperture measurements, we are left with a possible synthesis of both cases: the PBS is tied to the presence of icy grains, but the grains themselves trace chemical evolution of the outermost layers of the nucleus as it sheds material.
The inbound icy grains could be chemically distinct and produce the PBS, while the outbound (presumably pristine) grains observed by \citet{Protopapa_2018} would not necessarily exhibit the same activity.

Over the long storage lifetime of Oort-cloud comets there is significant chemical processing to a depth of 10 meters or more \citep{Garrod_2019, Gronoff_Maggiolo_2020, Maggiolo_2020}.
UV processing dominates the outer micron of the surface that occurs with a short ($\sim$ 1000 year) time scale, while galactic cosmic rays penetrate to tens of meters and slowly build up populations of complex organic molecules (COMs) and other notable chemical species like O$_2$, OH, and H$_2$O$_2$ \citep{Willis_2024}.
In the context of outer solar system bodies, \citet{Johnson_Quickenden_1997} show spectral evidence of trapped OH in the ices of Europa while pointing out that detrapping of OH in water ice at the $100 - 115$ K range is well-established, which corresponds to the surface temperature of an icy body at 7~AU \citep{Meech_Svoren_2004} (though thermal propogation to inner layers will lag behind).
The thin outer shell ($<10$ m) of nuclear material of dynamically new comets should be both chemically rich and distinct to the pristine inner material after long periods of radiolytic processing \citep{Meech_Svoren_2004}.
Further, an array of chemical populations can be trapped at low temperatures in both the highly porous nuclear matrix and larger voids, which can later be released directly from the nucleus as it warms, or detrapped as icy grains sublimate.

This picture predicts two major features.
The first is release of relatively slow-moving fragments we would otherwise expect to be high velocity, like the OH and CS produced via photodissociation from parent molecules H$_2$O and CS$_2$ respectively.
The second is chemical evolution over the apparition as the outer shell is lost to sublimation, which could account for the disappearance of the PBS post-perihelion as the radiolytically-processed material is exhausted.
\citet{Hyland_2019} derived spatial profiles of OH, NH, CN, C$_2$, and C$_3$ in dynamically new comet C/2014 S2 via two-dimensional spectra, showing simultaneous near-nucleus enhancements in column density of OH, NH, and C$_2$ extending to roughly $1 \times 10^4$ km projected distance.
These measurements were taken at 2.4 AU outbound after a perihelion at 2.08 AU, contrasted with the 0.82 AU perihelion of US$_{10}$: the lower rate of evaporation of material from C/2014 S2 would suggest an extended period of PBS activity if this behavior is linked to the outer shell of material, so these PBS-like features are consistent with the proposed mechanism despite occurring post-perihelion.
Taken together, the enhancements of these species point again toward a population of slower-moving molecules that are not consistent with the standard Haser and vectorial models of the coma.

\section{Conclusion}
We have developed a methodology for reducing and analyzing \textit{Swift} UVOT comet images over a broad range of heliocentric distances.
A central feature of the pipeline is its Bayesian treatment of the dust spectral gradient, \(S'\), which is treated as a source of uncertainty rather than as a fixed estimated value.
We adopt both aperture photometry and, when possible, radial profile extraction to derive OH and water production rates, along with other characteristics of interest such as $Af\rho$ and active area estimates.

We applied this pipeline to C/2013 US$_{10}$ (Catalina), a dynamically new long-period comet observed over a wide range of inbound and outbound heliocentric distances, achieving good agreement with previously established values of water production, $Af\rho$, and magnitude.
The analysis confirms OH detection at large pre-perihelion distance, followed by a substantial post-perihelion decline and non-detections by \(\sim 4\) AU outbound.
When interpreted with standard Haser or vectorial models, the large-\(r_h\) detections imply high water-production rates and large active areas, while the post-perihelion active area is closer to a steady value of order \(\sim 50~\mathrm{km^2}\).
This asymmetry suggests that the dominant source of OH changed over the apparition and that the large-\(r_h\) signal is difficult to interpret as ordinary nuclear sublimation alone on thermodynamic grounds.

Surface-brightness profiles reveal an additional compact excess in the UVW1 filter near the nucleus, which we refer to as the pale blue spot (PBS).
The feature is spatially coherent, evolves with time, and is not naturally reproduced by plausible dust-color variations, reduction artifacts, or the expected brightness and morphology of known UVW1 contaminants such as CO\(_2^+\), CO\(^+\), CS, CN, or atomic metal emission.
Although our data do not uniquely identify the carrier of the PBS, they support the conclusion that the feature is a real inner-coma component rather than an obvious artifact or contaminant.

Taken together, the large-\(r_h\) OH detections and the PBS suggest that the coma of US$_{10}$ may contain more than one OH-producing component.
One component is the canonical nuclear source associated with water sublimation and subsequent photodissociation, whose column density is predicted by Haser or vectorial models.
The second appears to be a compact, near-nucleus component that is most prominent pre-perihelion and persists only briefly after perihelion.
If the PBS signal is in fact dominated by OH, it may trace a slow or localized OH-producing population that does not follow the standard vectorial picture of photodissociated H$_2$O producing OH with $\sim 1$ km/s velocity kicks.
This hypothetical source of slow-moving molecules remains a mystery, however, because the data do not identify the particular physical release mechanism, and we cannot rule out contamination in the UVW1 filter with absolute certainty.

C/2013 US$_{10}$ is a particularly useful test case because water-ice grains were independently detected by TRAPPIST in long-term monitoring, but their link to the presence of the PBS is shaky.
It is possible that the PBS is only associated with icy grains enriched in radiolytically processed or otherwise chemically altered material from the outer layers of the nucleus.
In this picture, the near-nucleus excess is strongest before perihelion because the comet is still losing processed or compositionally distinct mantle material, and it weakens after perihelion as that reservoir is depleted through mass loss or thermally altered.
While speculative, this explanation provides a plausible connection between the transient PBS, the independently observed icy grains, and the evolutionary state of the nucleus.

If correct, such a mechanism may be most relevant for dynamically new comets, whose outer layers can preserve the cumulative effects of long-term irradiation and radiolytic processing before their first passages through the inner Solar System.
The disappearance of the PBS after perihelion, despite evidence for outbound icy grains, would then suggest that the relevant source is not simply ice itself, but a particular processed or volatile-rich component of the near-surface material that is preferentially exposed or exhausted during the early stages of activity.
This possibility emphasizes the value of systematic UV monitoring across multiple activity regimes: observations before, near, and after perihelion can provide necessary interpretational context needed to tease apart physical storage and release mechanisms, which directly inform models and alter production estimates.
Accurately characterizing volatile release may also relieve tension between scattered water production values measured with varying aperture sizes and model assumptions that should otherwise agree.
Future observations combining UV imaging, spectroscopy, and contemporaneous infrared constraints on icy grains will be needed to determine whether PBS-like features are common and to identify the mechanism responsible for the localized emission, as well as determine what other chemical species, if any, that the PBS might contain.

\begin{acknowledgments}
We thank Michael Siegel and the team at the Swift Mission Operations Center for their scheduling and running the observation campaign of comet US$_10$. We would like to thank the Swift project for awarding target of opportunity time for the monitoring of the comet. DB received support through NASA's Swift Guest Investigator program (Swift Cycle 12; NNX16AN70G). DB and JWN acknowledge support from NASA's ULTRASAT program (award number 80NSSC24K0383).
\end{acknowledgments}

\begin{contribution}
SO was authored the pipeline sotfware, performed the data reduction, and was responsible for writing and submitting the manuscript.
ZX edited the manuscript and contributed valuable expertise in \textit{Swift}/UVOT image processing and coma modeling.
JWM reviewed the manuscript and contributed interpretive context of cometary UV observations.
SJB reviewed the manuscript and contributed spectroscopic modeling.
DB scheduled the observations, edited the manuscript, and supervised the direction and development of the work described in the manuscript.
\end{contribution}

\facilities{\textit{Swift}(UVOT)}

\software{astropy \citep{astropy_2022},
          numpy \citep{numpy_harris2020},
          scipy \citep{scipy_virtanen_2020},
          sbpy \citep{Mommert2019},
          matplotlib \citep{matplotlib_hunter_2007},
          marimo \citep{marimo_Agrawal_2023}
          }


\bibliography{us10bib}{}
\bibliographystyle{aasjournalv7}



\end{document}